\documentclass[twocolumn,longbibliography]{revtex4-2}

\usepackage{hyperref}
\usepackage{graphicx}
\usepackage{epsf}
\usepackage{amsmath,amssymb}
\usepackage{color}

\def\Ni#1#2{{}^{#1#2}\textrm{Ni}}
\def\Ca#1#2{{}^{#1#2}\textrm{Ca}}
\def\Ti#1#2{{}^{#1#2}\textrm{Ti}}

\def\Si#1#2{{}^{#1#2}\textrm{Si}}
\def\S#1#2{{}^{#1#2}\textrm{S}}
\def\SA{{}^{A}\textrm{S}}
\def\NiA{{}^{A}\textrm{Ni}}
\def\Ti#1#2{{}^{#1#2}\textrm{Ti}}

\def\Ni#1#2{{}^{#1#2}\textrm{Ni}}

\usepackage{ulem}
\begin{document}
\preprint{KUNS-2877}
\title{Deformation effects on the surface neutron densities of stable S and Ni isotopes
probed by proton elastic scattering via isotopic analysis}

\author{Yoshiko Kanada-En'yo}
\affiliation{Department of Physics, Kyoto University, Kyoto 606-8502, Japan}

\begin{abstract}
To extract structure information from proton elastic scattering off S isotopes
at 320~MeV and Ni isotopes at $E_p=180$~MeV,
this study proposes isotopic analyses
combining nuclear structure and reaction calculations.
The isotonic analysis was repeated on $\Ca48$ and $\Ti50$.
The structure calculations were performed by using the spherical and deformed
Relativistic Hartree--Bogoliubov~(RHB) calculations, and the spherical
nonrelativistic Skyrme Hartree--Fock--Bogoliubov calculations.
The $(p,p)$ reactions were calculated
using the relativistic impulse approximation (RIA) assuming the theoretical densities of target nuclei.
The RIA calculations using the target densities obtained by the RHB calculation, along with the
density-dependent point-coupling interactions reasonably reproduced the $(p,p)$ cross sections in the studied mass number region.
The nuclear structure and $(p,p)$ reactions
were analyzed in detail; especially, the effects of deformation
on the isotopic systematics of
the surface densities and $(p,p)$ cross sections were clarified.
The deformation effects were found to be essential to describe the isotopic systematics of the experimental $(p,p)$
cross sections.
Overall, isotopic and isotonic analyses are useful for extracting the nuclear structure information
such as nuclear deformations and single-particle features from proton scattering,
enabling sensitive probing of surface densities.

\end{abstract}

\maketitle

\section{Introduction}
As a sensitive probe of surface neutron densities, proton elastic scattering
has been utilized for extracting detailed density profiles
determining the neutron skin thickness
of nuclei. Accordingly, experimental researches of $(p,p)$ reactions of various nuclei at intermediate energies
have been performed~\cite{Ray:1978ws,Ray:1979qv,Hoffmann:1980kg,Terashima:2008zza,Zenihiro:2010zz,Zenihiro:2018rmz}.
Reaction analyses under the relativistic impulse approximation (RIA) have precisely determined
the neutron densities from 295~MeV proton scattering data. These analyses employ
density-dependent effective $NN$ interactions \cite{Sakaguchi:1998zz} constructed from
the original Murdock and Horowitz~\cite{Horowitz:1985tw,Murdock:1986fs,RIAcode:1991}~(MH) model,
hereafter called the ``ddMH model''.
The RIA+ddMH model has successfully described 295~MeV $(p,p)$ reactions of various target nuclei, including
Sn~\cite{Terashima:2008zza}, Pb~\cite{Zenihiro:2010zz}, and Ca~\cite{Zenihiro:2018rmz} isotopes.

The $(p,p)$ cross sections sensitively probe the surface neutron density,
which is affected by structure properties
such as neutron single-particle properties and nuclear deformations.
Therefore, the $(p,p)$ reactions provide useful information on these properties.
A major advantage of proton elastic scattering is its higher quality data than
other reactions including inelastic scattering, knock-out, and pick-up reactions.
In my previous papers~\cite{Kanada-Enyo:2021oee,Kanada-Enyo:2021gqt}, I analyzed
the $\textrm{Pb}(p,p)$ and $\textrm{Sn}(p,p)$ reactions at 295~MeV, and showed that the cross sections were sensitive to
single-particle occupation of the low-$\ell$ orbit in the major shell.
For precise analyses with less model ambiguity,
I then proposed an isotopic analysis method combining structure and reaction calculations
for precise analyses with low model ambiguity and proved its
applicability.

In the present work, I present isotopic analyses of
proton elastic scattering from stable S and Ni isotopes
at the incident energies $E_p=180\sim 320$~MeV.
Apart from the magic nuclei, these nuclei
are med-shell nuclei and considered as deformed nuclei based on their
$E2$ transition properties determined by $\gamma$ rays and electron and hadron inelastic scattering data.
I perform structure calculations based on the spherical and deformed
Relativistic Hartree--Bogoliubov~(RHB) calculations~\cite{Niksic:2014dra}, and (for comparison)
the spherical nonrelativistic Skyrme Hartree--Fock--Bogoliubov~(SHFB)~\cite{Bennaceur:2005mx} calculations.
The $(p,p)$ reactions are calculated
by the RIA-ddMH model assuming the theoretical densities of target nuclei.
In particular, this study explores the deformation effects on isotopic systematics of
the surface densities and $(p,p)$ cross sections and
thus examines the sensitivity of the
$(p,p)$ cross sections to the structure properties.
The present application to stable nuclei aims to demonstrate
that isotopic analyses of $(p,p)$ cross sections at intermediate energies
can extract the structure information, including nuclear deformations
and single-particle features from nuclear reactions.
The isotonic analysis is repeated on $\Ca48$ and $\Ti50$ at $N=28$
to discuss the deformation
effects of $\Ti50$.

The remainder of this paper is organized as follows.
The structure and reaction calculations
are explained in Sec.~\ref{sec:calculations}, and the results of S isotopes, Ni isotopes, and $N=28$ isotones
are presented in subsections~\ref{sec:results-s}, ~\ref{sec:results-ni}, and \ref{sec:results-ti}, respectively.
A summary is given in Sec.~\ref{sec:summary}.

\section{Methods of nuclear structure and reaction calculations}\label{sec:calculations}

\subsection{Structure calculations}

To calculate the structures of the even--even nuclei,
the spherical and deformed RHB calculations were executed in
computational DIRHB code~\cite{Niksic:2014dra} with the
density-dependent point-coupling~(DD-PC1)~(pc1)~\cite{Niksic:2008vp}
and density-dependent meson-exchange (DD-ME2)~(me2)~\cite{Lalazissis:2005de} interactions.
In addition, the spherical SHFB
was executed in HFBRAD code~\cite{Bennaceur:2005mx} with
SKM*~\cite{Bartel:1982ed} and SLy4~\cite{Chabanat:1997un} interactions employing
mixed-type pairing forces.

\subsection{Calculations of proton elastic scattering reactions}

The proton elastic scattering at $E_p=180\sim 320$~MeV
was calculated using the RIA+ddMH model~\cite{Sakaguchi:1998zz}.
The nucleon--nucleus potentials were obtained by folding
the vector and scalar densities of target nuclei obtained by the structure calculations
as described in my previous papers~\cite{Kanada-Enyo:2021oee,Kanada-Enyo:2021gqt}.
The neutron (proton) vector densities were assumed as
the theoretical neutron (proton) densities, $\rho_n(r)$ ($\rho_p(r)$) and
the neutron (proton) scalar densities were set to
$0.96\rho_n(r)$ ($0.96\rho_p(r)$).
The same scalar densities were used in
analyses of the $\textrm{Sn}(p,p)$ and $\textrm{Pb}(p,p)$ reactions
(see Refs.~\cite{Terashima:2008zza,Zenihiro:2010zz}).

The effective $NN$ interactions in the RIA+ddMH model are based on the
meson-exchange model, which includes the density dependences of the
$\sigma$- and $\omega$-meson masses and coupling constants~\cite{Sakaguchi:1998zz,Terashima:2008zza,Zenihiro:2010zz}. These density dependencies are extensions of
the density-independent interactions in the original
Murdock and Horowitz model~\cite{Horowitz:1985tw,Murdock:1986fs,RIAcode:1991}.
The present RIA+ddMH calculations of
the $(p,p)$ reactions at $E_p=180\sim 320$~MeV adopt
the latest parameterization, which has been calibrated
to fit the $^{58}\textrm{Ni}(p,p)$ data at $295~$MeV~\cite{Zenihiro:2010zz}.

\section{Results of S isotopes} \label{sec:results-s}
\subsection{Structure properties around $\S32$}

%%%%%%%%%%%%%%%%%%%%%%%%%%%%%%
\begin{figure}[!htpb]
\includegraphics[width=0.5\textwidth]{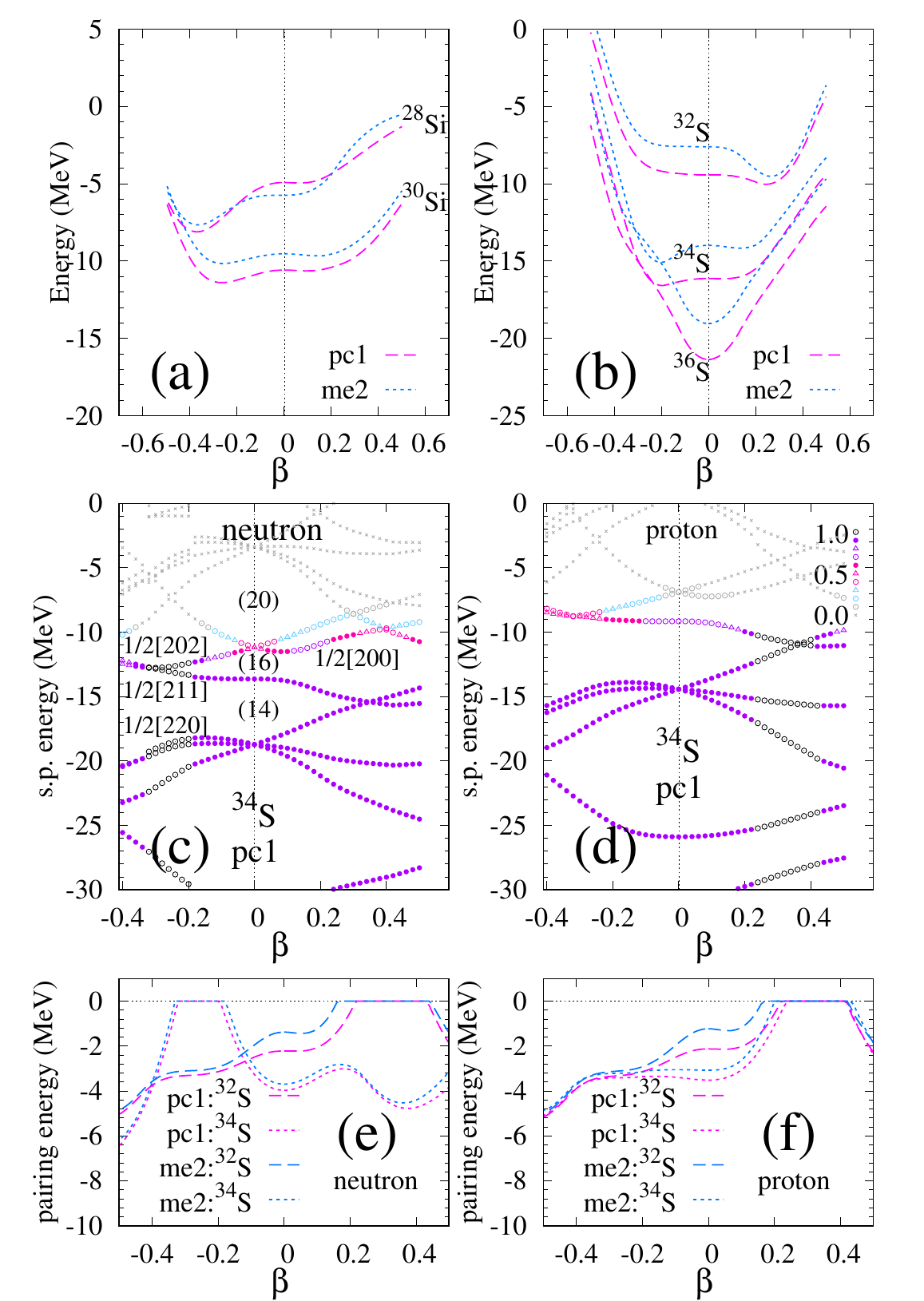}
\caption{Energy curves, single-particle energies, and pairing energies of Si and S isotopes
obtained by the deformed RHB calculations with the pc1 and me2 interactions.
Energy curves of (a) Si isotopes and (b) S isotopes;
(c) neutron and (d) proton
single-particle energies in $\S34$ calculated using the pc1 interactions
(symbol colors indicate the occupation probabilities).
(e) Neutron and (f) proton pairing energies in $\S32,$ and $\S34$.
\label{fig:enesp-s34}}
\end{figure}
%%%%%%%%%%%%%%%%%%%%%%%%%

%%%%%%%%%%%%%%%%%%%%%%%%%%%%%%
\begin{figure}[!htpb]
\includegraphics[width=0.5\textwidth]{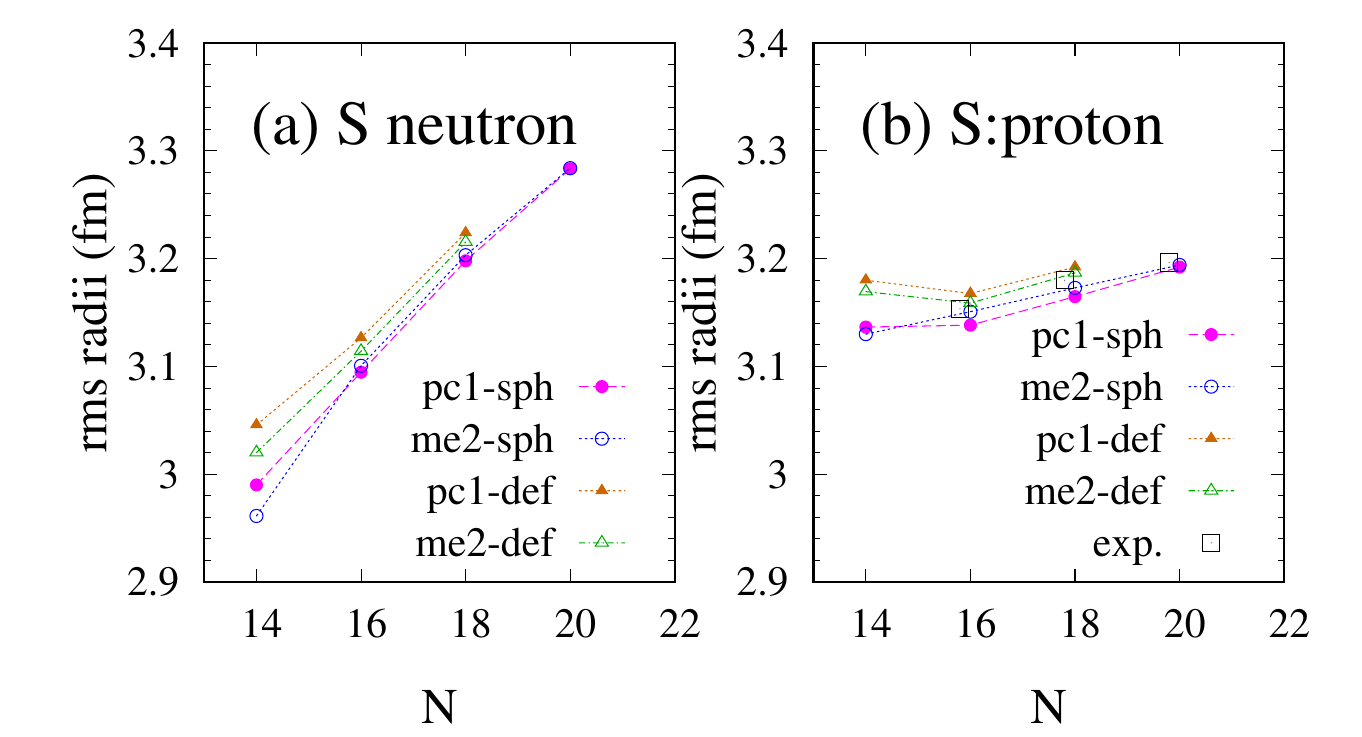}
\caption{Root-mean-square~(rms) (a)neutron and (b) proton radii of S isotopes in
the pc1-def and me2-def states for the deformed cases and in the
pc1-sph and me2-sph states for the spherical cases obtained by RHB calculations.
Also plotted are the
experimental rms proton
radii evaluated from the rms charge radii \cite{Angeli:2013epw}.
\label{fig:rmsr-s}}
\end{figure}
%%%%%%%%%%%%%%%%%%%%%%%%%

The structures of the Si and S isotopes were calculated by the spherical and deformed RHB calculations
with pc1 and me2 interactions.
Figure~\ref{fig:enesp-s34} shows the
$\beta$ dependences of the energies of Si and S isotopes (panels (a) and (b), respectively),
the neutron and proton single-particle energies and occupation probabilities in $\S34$ calculated using
the pc1 interaction (panels (c) and (d), respectively),
and the pairing energies in $\S32$ and $\S34$ (panels (e) and (f), respectively).
As shown in the energy curves, the Si and S isotopes around $\S32$ are mid-shell nuclei that favor nuclear deformation.
At the energy minima,
$\Si28$ and $\Si30$ are oblately deformed because by the $Z=14$ shell effect,
whereas the deformations of $\S32$, $\S34$, and $\S36$
change from prolate to oblate to oblate as the neutron number increases.
The prolate deformation of $\S32$ is due to the single-particle energy gap between $1/2[211]$ and
$1/2[200]$ orbits, whereas the oblate deformation of $\S34$ can be attributed to
the energy gap between the neutron $1/2[202]$
and $1/2[200]$ orbits. $\S36$ is spherical due to the $N=20$ shell effect.
The minimum-energy states of S and Si isotopes obtained by the deformed pc1(me2) calculations
are here labeled
``pc-def(me2-def)'' and the spherical states at $\beta=0$ are called ``pc-sph(me2-sph)''.

The neutron and proton radii of S isotopes in the pc1-def, me2-def, pc1-sph, and me2-sph
states are presented in Fig.~\ref{fig:rmsr-s}, together with the experimental proton radii obtained
from the charge radii.
The spherical states yield slightly smaller radii than the
deformed states, but both
states depend similarly on $N$.
The theoretical proton radii of the deformed and spherical states
reasonably agree with the experimental data of $\S32$, $\S34$, and $\S36$.

\begin{figure}[!h]
\includegraphics[width=0.5\textwidth]{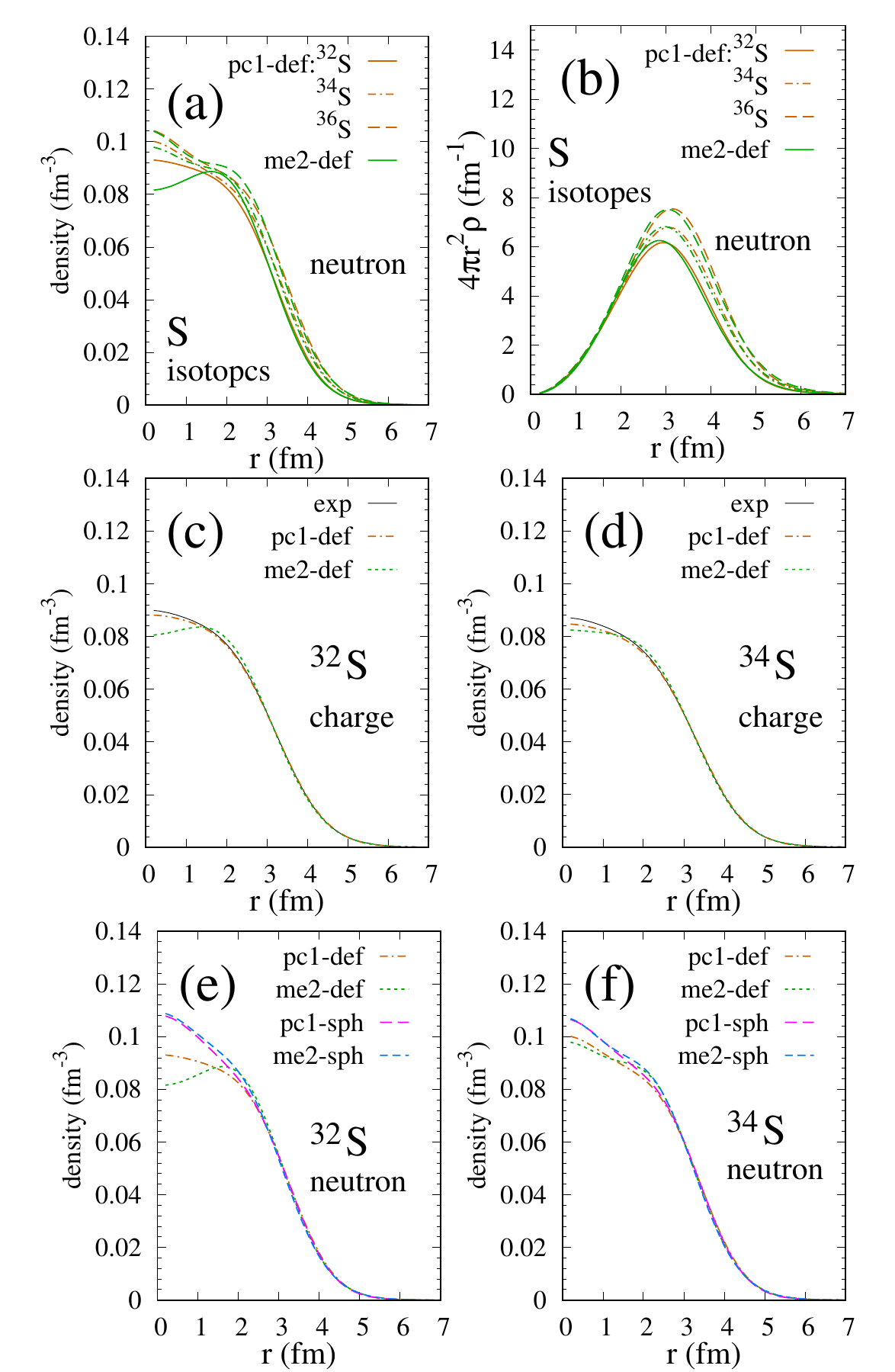}
\caption{Densities of S isotopes obtained in the deformed pc1 and me2 calculations:
(a) neutron densities $\rho_n(r)$ and (b) $4\pi r^2\rho_n(r)$
of S isotopes, charge densities of (c) $\S32$ and (d) $\S34$ in the pc1-def and me2-def states,
together with the experimental charge densities; neutron densities of
(e) $\S32$ and (f) $\S34$ for the pc1-def, me2-def, pc1-sph, and me2-sph states.
The experimental charge densities
were obtained from the sum-of-Gaussian~(SOG) fitting parameters listed in Ref.~\cite{DeJager:1987qc}.
\label{fig:dens-s}}
\end{figure}

The calculated neutron densities $\rho_n$ of the pc1-def and me2-def states of
S isotopes are presented in Fig.~\ref{fig:dens-s}(a) and the calculated $4\pi r^2\rho_n(r)$ are drawn in
Fig.~\ref{fig:dens-s}(b).
Panels (c) and (d) of this figure display the theoretical and experimental charge densities
of $\rho_\textrm{ch}$ of $\S32$ and $\S34$, respectively.
The pc1-def result well agrees with experimental charge densities
of the S isotopes.

Panels (e) and (f) of Fig.~\ref{fig:dens-s} compare
the neutron densities of the deformed and spherical states.
The internal neutron densities of the pc1-def, me2-def, pc1-sph, and me2-sph
scenarios visibly differ in the $r \lesssim 2$ region.
In $\S32$,
the internal neutron densities are enhanced in the spherical states
pc1-sph and me2-sph, where there is significant $1s_{1/2}$ occupation,
but are quenched in the deformed states pc1-def and me2-def
because prolate deformation decreases the $1s_{1/2}$ occupation.
In the deformed case,
the $1s_{1/2}$ component mixes in the $1/2[200]$ orbit above the Fermi level.
The central neutron density is lowest in the me2-def state
because the pairing gap vanishes in deformed $\S32$ at $\beta=0.20$.
These single-particle properties
contribute to the surface neutron densities in $r \gtrsim 3$~fm,
although the surface neutron densities of the deformed and spherical states show no obvious differences
in the figures.
In $\S34$, the difference in central neutron density
between the spherical and deformed states is smaller than in $\S32$.
Note that the proton scattering at $E\sim 300$~MeV is sensitive to
surface neutron density but not to internal density.

%%%%%%%%%%%%%%%%%%%%%%%%%%%%%%
%%%%%%%%%%%%%%%%%%%%%%%%%

\subsubsection{Proton scattering and isotopic analysis of S isotopes}

%%%%%%%%%%%%%%%%%%%%%%%%%%%%%%
\begin{figure}[!htpb]
\includegraphics[width=8 cm]{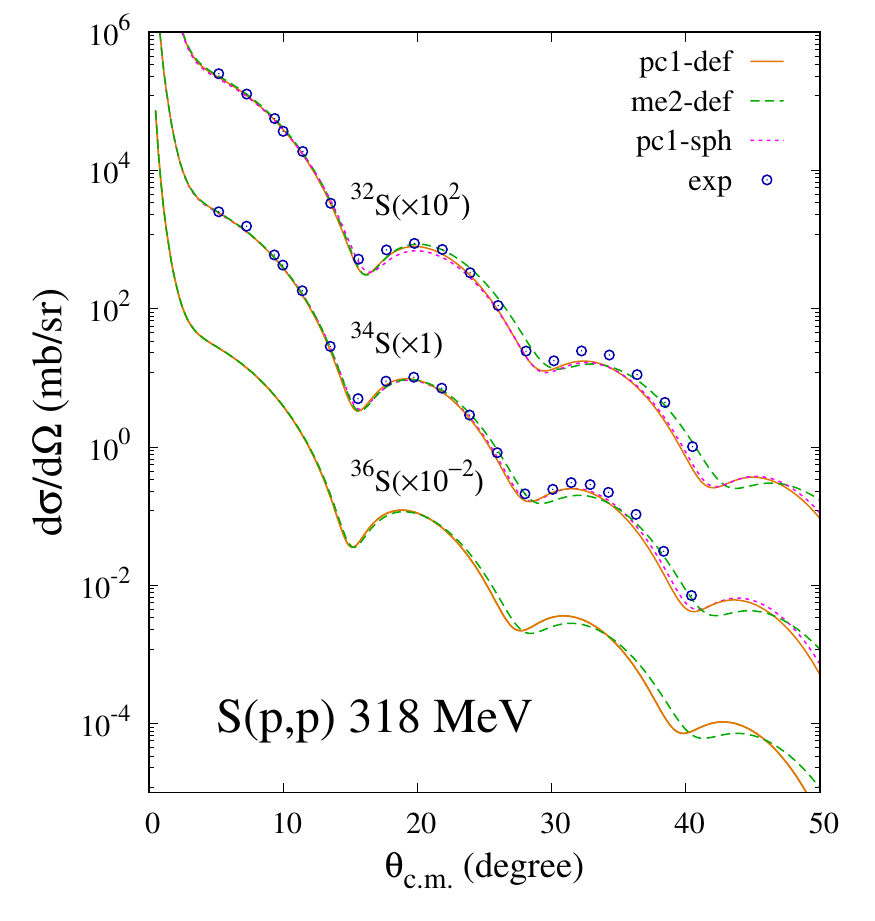}
\caption{$\SA(p,p)$ cross sections at 318~MeV calculated by the RIA-ddMH employing the pc1-def, me2-def, and pc1-sph densities
of S isotopes.
The experimental cross sections from Ref.~\cite{Kelly:1991zza}
are plotted for comparison.
\label{fig:cross-s}}
\end{figure}
%%%%%%%%%%%%%%%%%%%%%%%%%

%%%%%%%%%%%%%%%%%%%%%%%%%%%%%%
\begin{figure}[!htpb]
\includegraphics[width=5 cm]{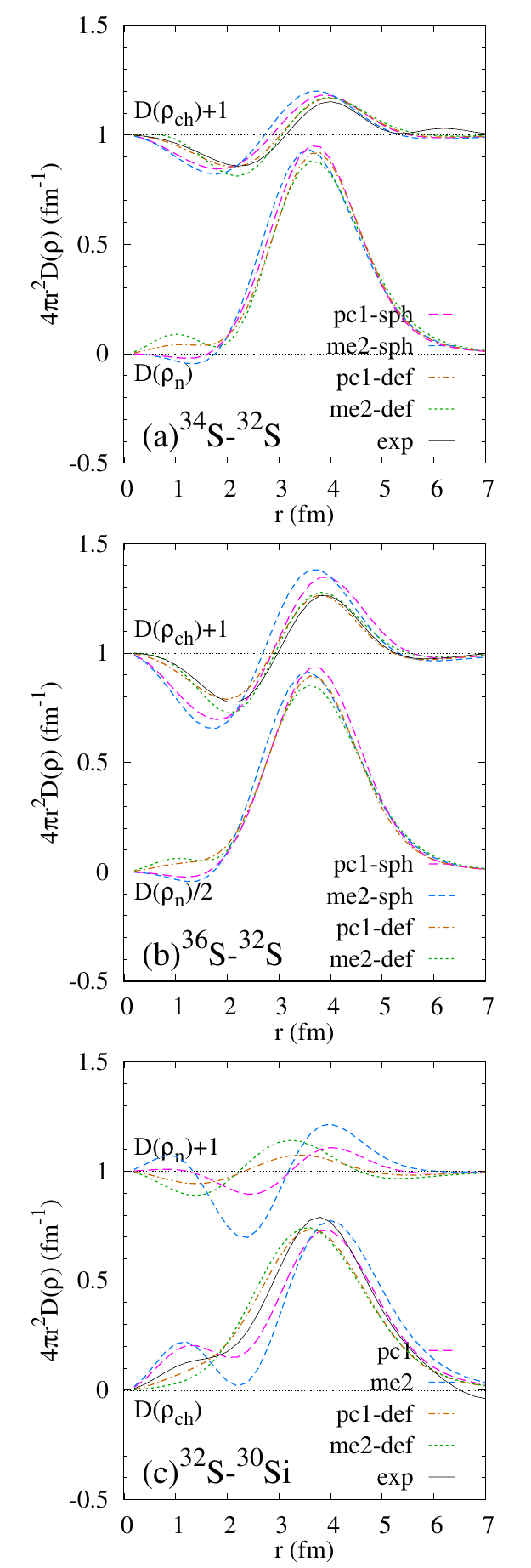}
\caption{
Neutron and charge density differences
in the pc1-def and me2-def states of S and Si isotopes
obtained by the deformed RHB calculations, compared with those of the
spherical states (pc1-sph and me2-sph):
(a) $\S34-\S32$
(b) $\S36-\S32$, and (c) $\S32-\Si30$.
The experimental charge density differences
obtained from Ref.~\cite{DeJager:1987qc} are also presented.
$4\pi r^2 D_n(r)$ and $4\pi r^2 D_\textrm{ch}(r)+1$ are plotted in panels (a) and (b), and
$4\pi r^2 D_\textrm{ch}(r)$ and $4\pi r^2 D_n(r)+1$ are plotted in panel (c).
\label{fig:dens-compare-s}}
\end{figure}
%%%%%%%%%%%%%%%%%%%%%%%%%

%%%%%%%%%%%%%%%%%%%%%%%%%%%%%%
\begin{figure}[!htpb]
\includegraphics[width=8 cm]{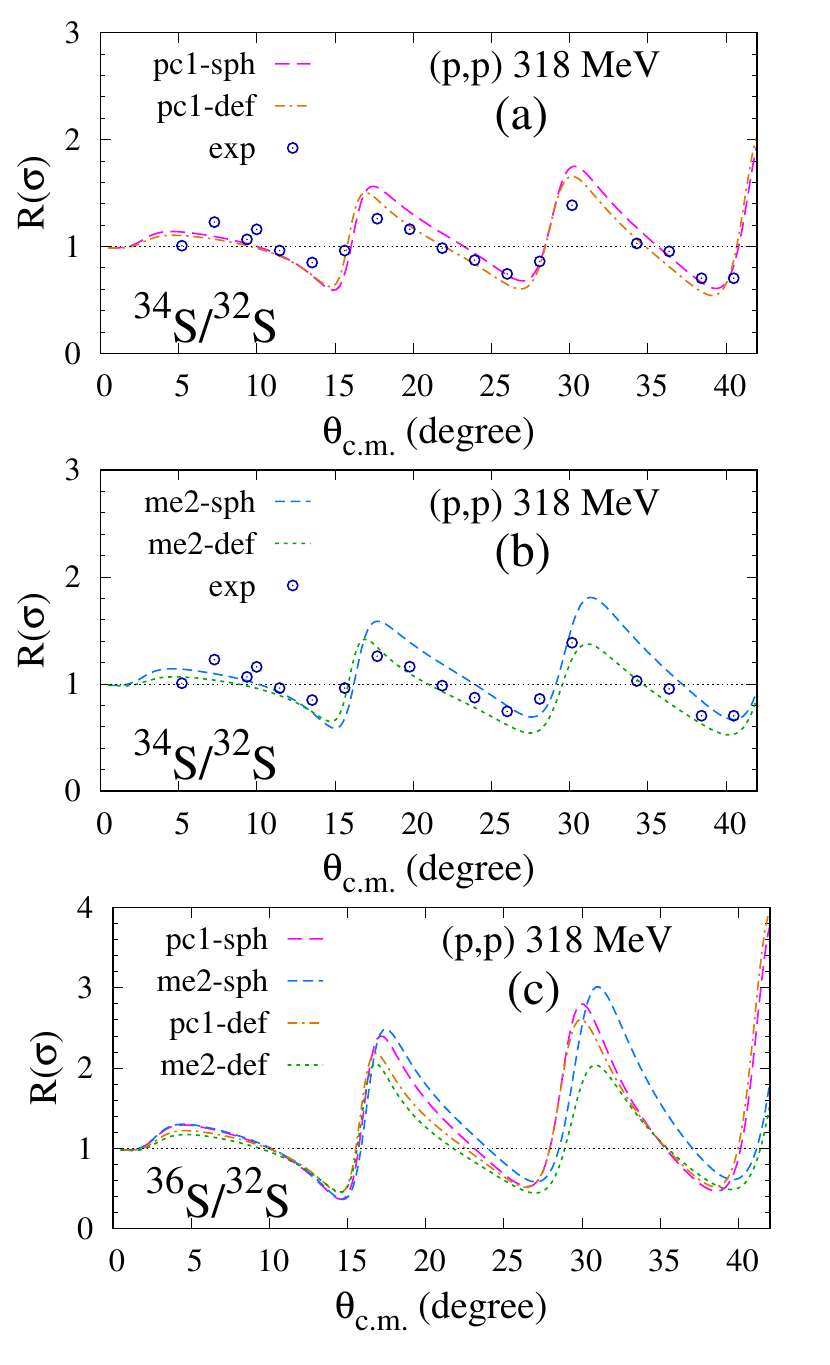}
\caption{
Isotopic cross section ratios $R(\Omega)$
of the $\S34/\S32$ ratios of the $\SA(p,p)$ reactions at
318~MeV obtained by the RIA-ddMH calculations employing
(a) pc1-def and pc1-sph and (b) me2-def and me2-sph densities.
The experimental values obtained from the cross section data
of Ref.~\cite{Kelly:1991zza} are plotted for comparison.
(c) $R(\Omega)$ for the $\S36/\S32$ cross section ratios of 318~MeV
$(p,p)$ reaction.
\label{fig:cross-compare-s}}
\end{figure}
%%%%%%%%%%%%%%%%%%%%%%%%%

The $(p,p)$ reactions of $\S32$, $\S34$, and $\S36$ at 318~MeV were calculated within the RIA+ddMH model
employing the pc1-def and me2-def
densities of the deformed states obtained by the RHB calculations.
The calculated cross sections are compared with the experimental data
in Fig.~\ref{fig:cross-s}.
The pc1-def and me2-def results are consistent with the experimental data.
The results of the spherical calculations are qualitatively similar to the deformed
calculations.
To thoroughly investigate the deformation effect on the $(p,p)$ cross sections,
I perform isotopic analyses of the densities and cross sections of the S isotopes
and compare the results of the deformed and spherical states.

The isotopic density differences from those of $\S32$ are defined as
\begin{align}
&D(\rho_{n};r)\equiv \rho_{n}(\SA;r)-\rho_{n}(\S32;r),\\
&D(\rho_\textrm{ch};r)\equiv \rho_\textrm{ch}(\SA;r)-\rho_\textrm{ch}(\S32;r),
\end{align}
and the isotopic cross section ratios to $\S32$ are defined as
\begin{align}
R(\sigma;\theta_\textrm{c.m.})\equiv \frac{d\sigma(\SA)/d\Omega}{d\sigma(\S32)/d\Omega}.
\end{align}
The calculated isotopic density differences and
cross section ratios are compared with the experimental values
in Figs.~\ref{fig:dens-compare-s} and \ref{fig:cross-compare-s}, respectively.
The pc1-def and me2-def results of $4\pi r^2 D(\rho_\textrm{ch})$ for the deformed states
agree with the experimental $\S34-\S32$ and $\S36-\S32$ differences,
but the pc1-sph and me2-sph results for the spherical states disagree with the experimental differences.
In $4\pi r^2 D(\rho_n)$ plots of the neutron density, one can clearly distinguish the model dependence
of the surface neutron densities around $r \sim 3$~fm between deformed and spherical states.
$4\pi r^2 D(\rho_n)$ are slightly suppressed in the pc1-def and me2-def results of the
deformed states because the deformation affects the single-particle orbits and occupations
at the Fermi level, mainly in $\S32$ as described previously.
In the spherical states, the $1s_{1/2}$ orbit is almost occupied in $\S32$ and
the additional two neutrons in $\S34$ dominantly occupy the $0d_{3/2}$ orbit, so
$4\pi r^2 D(\rho_n)$ increases in the $r\sim 3$ region.
However, in the deformed case, the $1s_{1/2}$ component is mixed in the $1/2[200]$ orbit
and partially unoccupied in the prolate $\S32$ state. Therefore, the $1s_{1/2}$ component
is partially occupied by the additional two neutrons in the oblate $\S34$ state.
Consequently, the $0d_{3/2}$ contribution to the $\S34-\S32$ difference
slightly decreases and
$4\pi r^2 D(\rho_n)$ in the $r\sim 3$ region is decreased to some extent.
This deformation effect on the isotopic difference $4\pi r^2 D(\rho_n)$ of the surface neutron densities
can be sensitively probed by the $(p,p)$ cross sections in the isotopic cross section ratios $R(\sigma)$
presented in Fig. \ref{fig:cross-compare-s}.
In the calculations with the deformed states (pc1-def and me2-def),
the peak $R(\sigma)$ amplitudes are smaller than in the calculations with the spherical states (pc1-sph and me2-sph),
because $4\pi r^2 D(\rho_n)$ around $r \sim 3$~fm is smaller than in the spherical states.
The calculated $\S34/\S32$ cross section ratios more closely match
the experimental $R(\sigma)$ data in the pc1-def and me2-def states
than the pc1-sph and me2-sph states,
indicating that deformations of the $\S32$ and $\S34$ structures
are essential for describing the isotopic systematics of $(p,p)$ cross sections.

%%%%%%%%%%%%%%%%%%%%%%%%%%%%%%
%\begin{figure}[!h]
%\includegraphics[width=8 cm]{Ay-s-fig.eps}
% \caption{
%\label{fig:Ay-s}}
%\end{figure}
%%%%%%%%%%%%%%%%%%%%%%%%%

\section{Results of Ni isotopes around $N=32$} \label{sec:results-ni}
\subsection{Structure properties of Ni isotopes}

\begin{figure}[!h]
\includegraphics[width=0.5\textwidth]{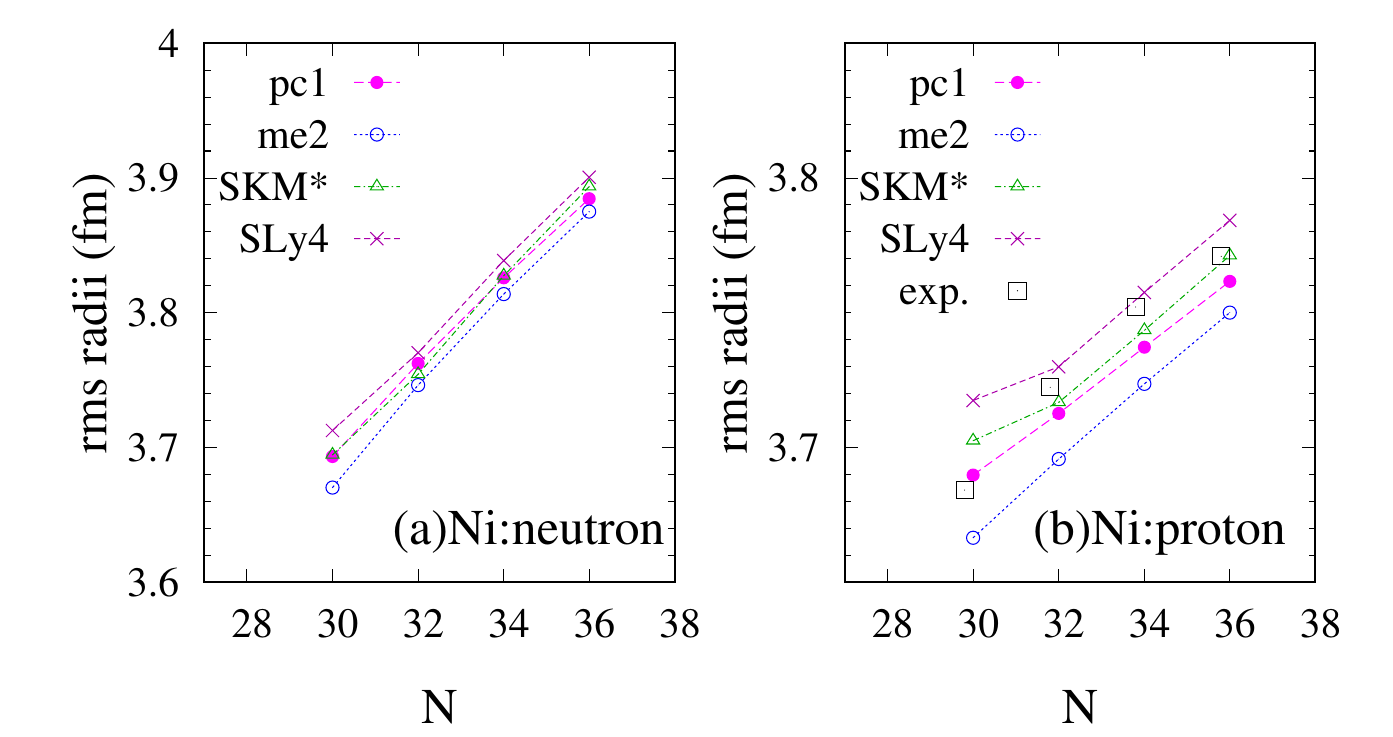}
\caption{
Root-mean-square~(rms) (a) neutron and (b) proton radii of Ni isotopes
obtained in the spherical calculations of the RHB with the pc1 and me2 interactions
and the SHFB with the SKM* and SLy4 interactions.
The experimental rms proton
radii evaluated from the rms charge radii \cite{Angeli:2013epw} are also presented.
\label{fig:rmsr-ni}}
\end{figure}

For structure calculations of Ni isotopes,
I implemented the spherical and deformed RHB models with the pc1 and me2
interactions and (for comparison) the spherical SHFB models with SKM* and SLy4 interactions.
The root-mean-square neutron $(r_n)$ and proton ($r_p$) radii obtained in the spherical RHB and SHFB calculations
are presented in Fig.~\ref{fig:rmsr-ni}.
The spherical pc1, me2, SKM*, and SLy4 calculations yield qualitatively similar $N$ dependences of $r_n$ and $r_p$.
The $N$ dependence of the calculated $r_p$ deviates
from the data.
In particular,
the SKM* and SLy4 results exhibit a kink at $N=32$ due to the shell effect,
whereas the $r_p$ from $\Ni58$($N=30$) to $\Ni60$($N=32$) is enhanced in the data.
The pc1 and me2 results present no kink because the $N=32$ shell effect is smeared by the pairing effect, but
they slightly underestimate the difference of $r_p$ between $N=30$ and $N=32$.
Nuclear deformation is necessary for describing the $N$ dependence of $r_p$ in Ni isotopes around $N=32$, as discussed later.

%%%%%%%%%%%%%%%%%%%%%%%%%%%%%%
%%%%%%%%%%%%%%%%%%%%%%%%%

%%%%%%%%%%%%%%%%%%%%%%%%%%%%%%
\begin{figure}[!htpb]
\includegraphics[width=0.5\textwidth]{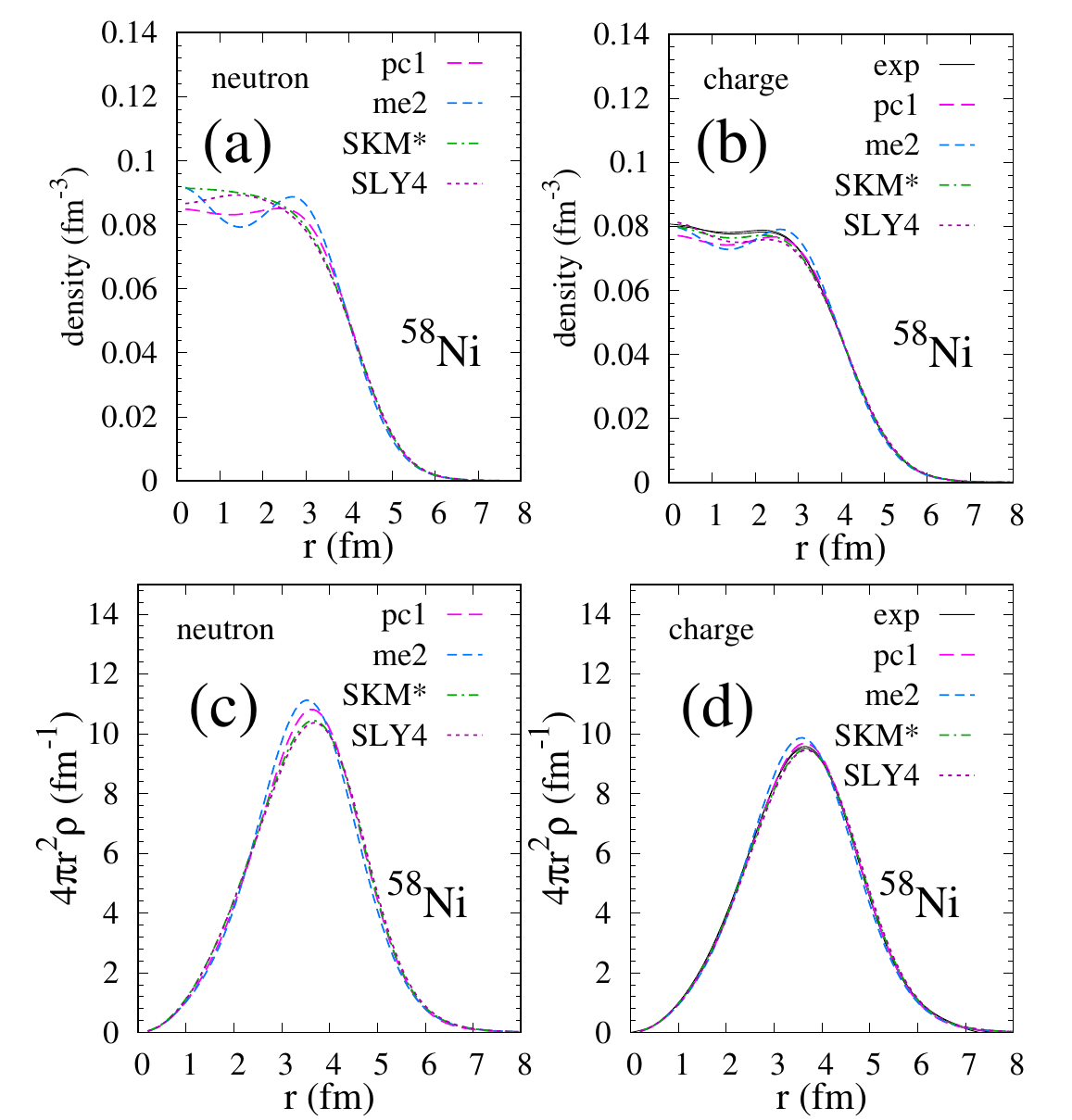}
\caption{
Radial distributions of $\Ni58$ densities obtained in the spherical pc1, me2, SKM*, and SLy4 calculations:
(a) neutron ($\rho_n$) and (b) charge ($\rho_\textrm{ch}$) densities;
(c) $4\pi r^2\rho_n(r)$; (d) $4\pi r^2\rho_\textrm{ch}$.
The experimental (exp) charge densities were
obtained from the sum-of-Gaussian~(SOG) fitting parameters listed in Ref.~\cite{DeJager:1987qc}.
\label{fig:dens-ni}}
\end{figure}
%%%%%%%%%%%%%%%%%%%%%%%%%

Figure~\ref{fig:dens-ni} compares the experimental neutron and charge densities of $\Ni58$ with those obtained in the spherical RHB and SHFB calculations.
The pc1, SKM*, and SLy4 calculations reasonably reproduce the experimental charge density at the nuclear surface in the
$r=3 \sim 5$~fm region.
The pc1 calculation best match the detailed profile
of $4\pi r^2\rho_\textrm{ch}$ around the surface peak.
In contrast, the me2 calculation fails to reproduce the surface charge density of $\Ni58$.

%%%%%%%%%%%%%%%%%%%%%%%%%%%%%%
\begin{figure}[!htpb]
\includegraphics[width=0.5\textwidth]{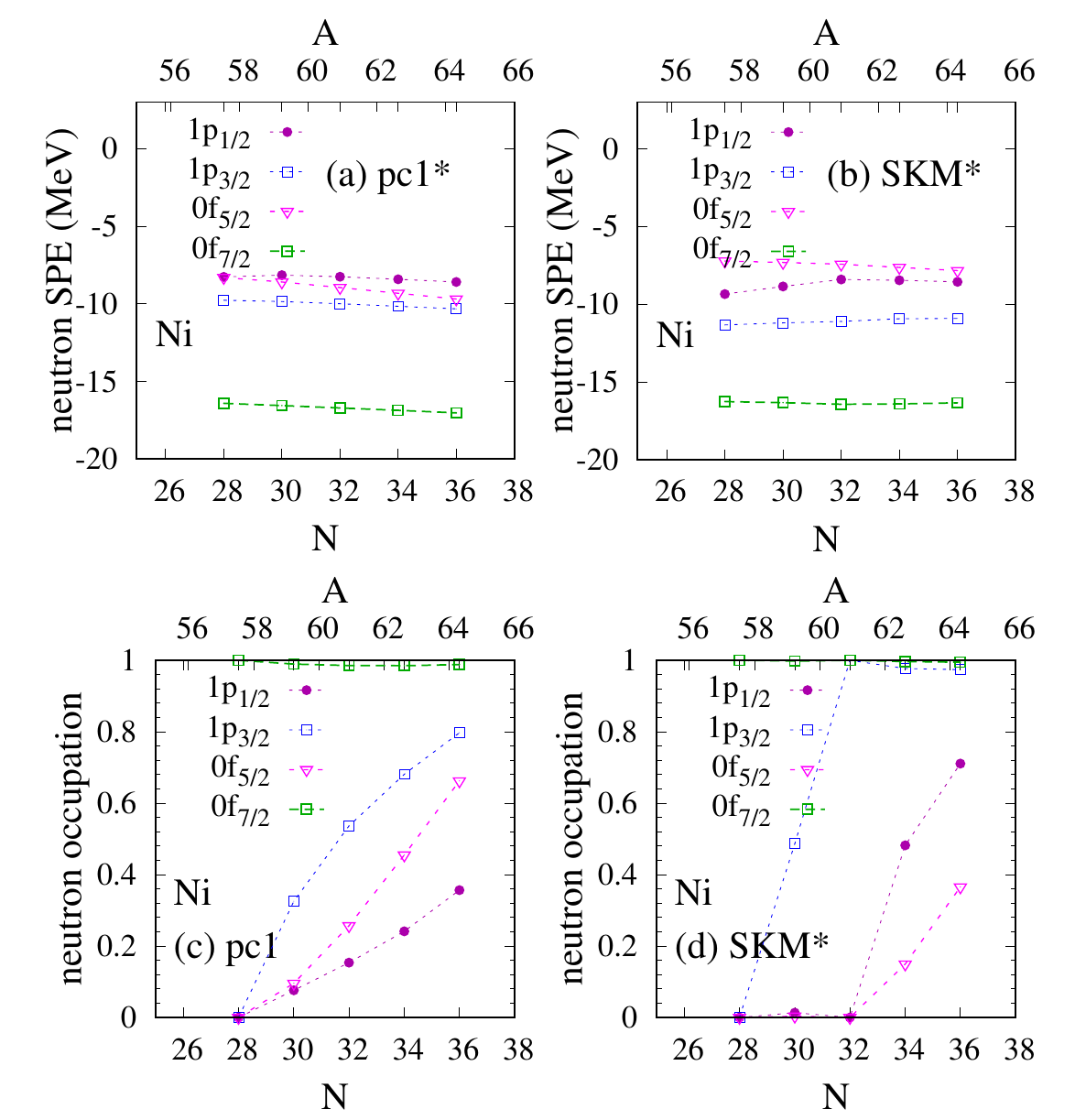}
\caption{
(a) (b) Neutron single-particle energies and (c) (d) occupation probabilities of Ni isotopes.
the results of the spherical pc1 calculations [panels (a) and (c)]
and the spherical SKM* calculations [panels (b) and (d)].
\label{fig:spspe-ni}}
\end{figure}
%%%%%%%%%%%%%%%%%%%%%%%%%

Figure~\ref{fig:spspe-ni} plots the neutron single-particle energies and occupation probabilities in the Ni isotopes obtained by the spherical pc1 and SKM* calculations.
In the pc1 result,
the major shell orbits $1p_{3/2}$, $0f_{5/2}$, and $1p_{1/2}$ are almost degenerate,
so the occupations of these orbits gradually increase along the isotope chain
from $\Ni56$ to $\Ni64$
due to the pairing effect.
The features of the single-particle properties are quite different in the SKM* results. The
$1p_{3/2}$-$0f_{5/2}$ energy gap in this calculation reached $1\sim 2$~MeV; accordingly,
the neutron pairing gap vanishes in $\Ni58$ and $\Ni60$.
This shell effect produces the kink of $r_p$ at $N=32$ in the SKM* result, as mentioned previously.
It should be commented that the $N=32$ closure in $\Ni60$ is inconsistent with the direct measurements
of spectroscopic factors in the $\Ni60(p,d)\Ni59$ reaction \cite{Matoba:1995sil}.

%%%%%%%%%%%%%%%%%%%%%%%%%%%%%%
\begin{figure}[!htpb]
\includegraphics[width=0.5\textwidth]{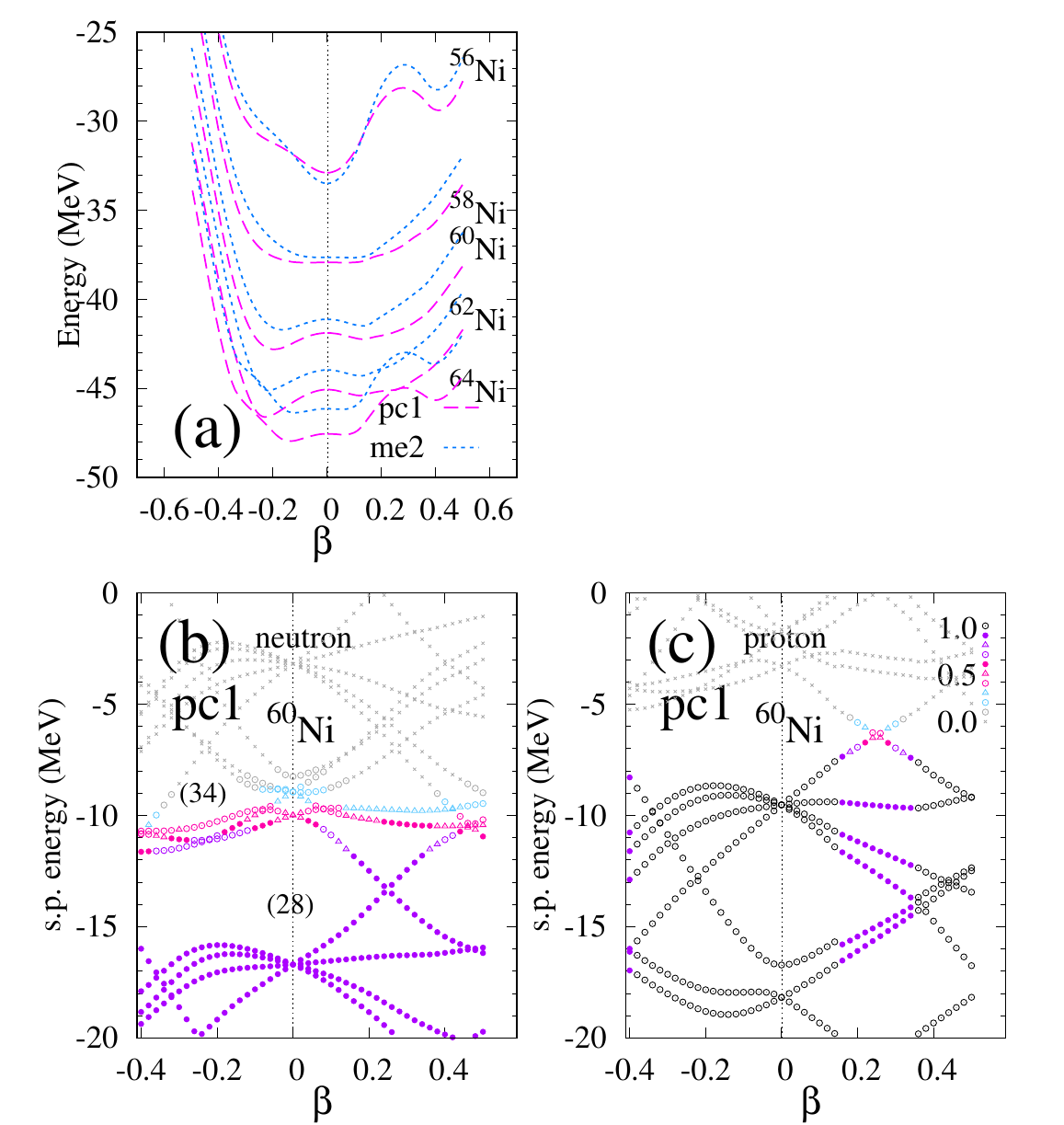}
\caption{
(a) Energy curves of Ni isotopes
obtained in the deformed RHB calculations using the pc1 and me2 interactions.
(b) neutron and (c) proton
single-particle energies in $\Ni60$ calculated using the pc1 interactions
(symbol colors indicate the occupation probabilities).
\label{fig:enesp-ni60}}
\end{figure}
%%%%%%%%%%%%%%%%%%%%%%%%%

Figure ~\ref{fig:enesp-ni60} shows the energy curves of
Ni isotopes ($A=56$--$64$) calculated by the deformed RHB model
and the SPE in $\Ni60$.
$\Ni56$ is a double magic nucleus and favors a spherical shape.
In contrast, $\Ni60$ and $\Ni62$ favor oblate shapes because an $N=34$ shell gap exists
at $\beta\sim -0.20$. The pc1 calculation yields minimum-energy states
of $\Ni60$(at $\beta=-0.20$) and $\Ni62$(at $\beta=-0.23$).
The energy curve of $\Ni58$ obtained by the pc1 calculation exhibits an intermediate feature.
The minimum-energy state of $\Ni58$(at $\beta=-0.08$) in this calculation shows a weak oblate deformation.
However, the energy curve is almost flat in the $\beta=-0.2\sim 0.1$ region
indicating $\beta$ softness of $\Ni58$.

\subsection{Proton scattering and isotopic analysis of Ni isotopes}

%%%%%%%%%%%%%%%%%%%%%%%%%%%%%%
\begin{figure}[!htpb]
\includegraphics[width=8cm]{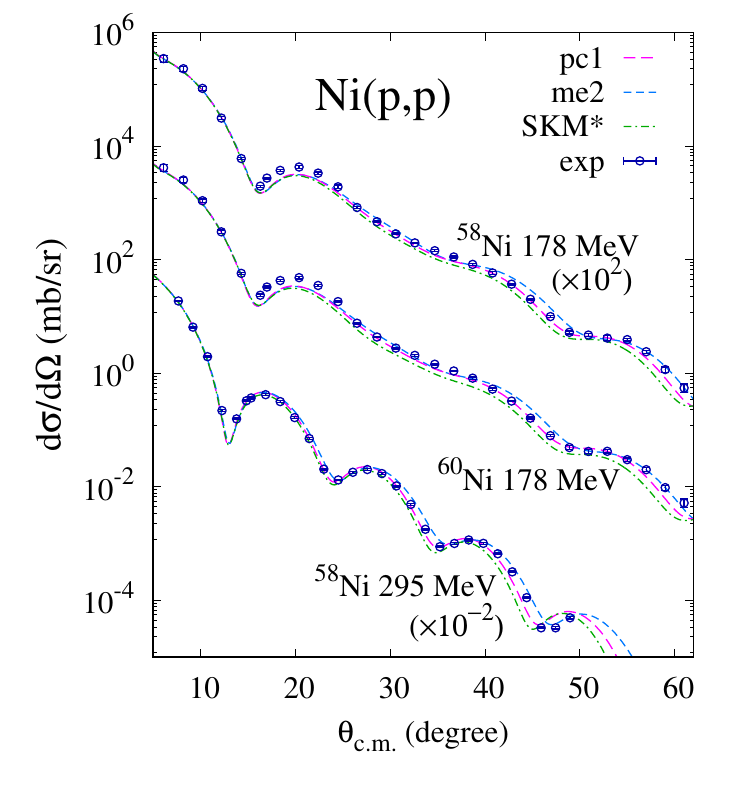}
\caption{
$\Ni58(p,p)$ and $\Ni60(p,p)$ cross sections at 178~MeV and $\Ni58(p,p)$ cross sections at 295~MeV
calculated by the RIA-ddMH model employing the Ni densities obtained in the spherical pc1, me2, and SKM*
calculations are compared with the experimental cross sections at 178~MeV~\cite{Ingemarsson:1981bm} and 295~MeV\cite{Zenihiro:2010zz}.
\label{fig:cross-ni}}
\end{figure}
%%%%%%%%%%%%%%%%%%%%%%%%%

%%%%%%%%%%%%%%%%%%%%%%%%%%%%%%
%\begin{figure}[!h]
%\includegraphics[width=8 cm]{Ay-ni-fig.eps}
% \caption{
%\label{fig:Ay-ni}}
%\end{figure}
%%%%%%%%%%%%%%%%%%%%%%%%%

The $(p,p)$ reactions
at $E_p=178$~MeV and 295~MeV
were calculated using the RIA+ddMH model with the theoretical Ni densities.
Figure~\ref{fig:cross-ni} presents
the cross sections obtained using
the spherical RHB calculations with pc1 and me2,
together with the results of the spherical SHFB calculations with
SKM* and the experimental cross sections.
The pc1 result reasonably agrees with the experimental cross sections of
$\Ni58(p,p)$ at $178$~MeV and 295~MeV and $\Ni60(p,p)$ at 178~MeV,
whereas the me2 and SKM* results deviate from the data,
especially at backward angles.

The$(p,p)$ cross sections calculated with the deformed Ni densities
The calculations with the deformed Ni densities
are qualitatively similar to those calculated with the spherical Ni densities.
To investigate the detailed deformation effects of Ni isotopes on the $(p,p)$ cross sections,
I compared the deformed and
spherical pc1 calculations in an isotopic analysis.
For the deformed states of Ni isotopes,
I chose the minimum-energy states obtained by the
deformed pc1 calculation, namely,
$\Ni58(\beta=-0.08)$, $\Ni60(\beta=-0.20)$, and $\Ni62(\beta=-0.23)$.
This default set was named
``pc1-def''.
Considering the $\beta$ softness of $\Ni58$,
I adopted an optional larger deformation of $\Ni58$ at $\beta=-0.17$.
This optional set $\{\Ni58(-0.17),\Ni60(-0.20),\Ni62(-0.23)\}$ was named ``pc1-def2''.
For comparison, the spherical states
at $\beta=0$ were collected into a dataset ``pc1-sph''.

\begin{figure}[!h]
\includegraphics[width=0.5\textwidth]{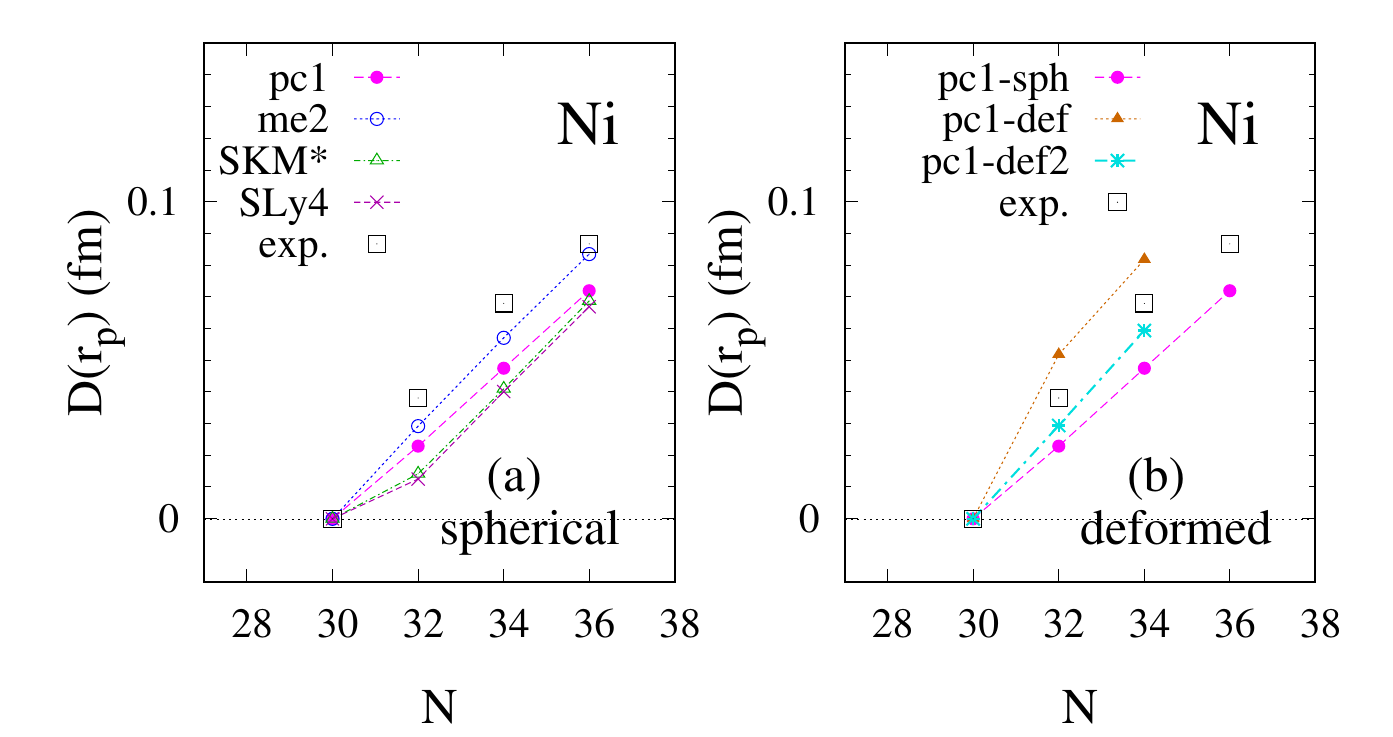}
\caption{
(a) Isotopic differences $D(r_p)\equiv r_p(\NiA)-r_p(\Ni58)$ in root-mean-square proton radii from
that of
$\Ni58$ obtained by the spherical calculations, and (b) $D(r_p)$ of the pc1-def, pc1-def2, and
pc1-sph states obtained by the deformed RHB calculations with the pc1 interaction.
The experimental (exp) data were evaluated from the charge radii \cite{Angeli:2013epw}.
\label{fig:rmsr-del-ni}}
\end{figure}

Panels (a) and (b) of Fig.~\ref{fig:rmsr-del-ni} plot the isotopic differences $D(r_p)\equiv r_p(\NiA)-r_p(\Ni58)$ of the $r_p$
in the spherical and deformed states, respectively,
together with the experimental values.
These results reveal the $N$ dependence of the proton radii $r_p$ measured from $r_p$ of .$\Ni58$.
The spherical calculations underestimate $D(r_p)$ of $\Ni60$ and $\Ni62$ [Fig.~\ref{fig:rmsr-del-ni}(a)].
The pc1-def, pc1-def2, and def-sph results of the deformed states are
presented in Fig.~\ref{fig:rmsr-del-ni}(b).
Owing to deformation effects, the
$D(r_p)$ is larger in the deformed pc results than in the spherical results.
In the pc1-def result, $r_p$ suddenly increases from $\Ni58(-0.08)$ to $\Ni60(-0.20)$ and the experimental
$D(r_p)$ is overestimated at $N=32$ and $N=34$. In the pc1-def2 results, where the deformation changes are modest
as $\Ni58(-0.17)$, $\Ni60(-0.20)$, and $\Ni62(-0.23)$ states,
the $D(r_p)$ values better match the experimental $D(r_p)$ at $N=32$ and 34 than the pc1-def and pc1-sph results.

To investigate the deformation effects on the isotopic systematics of the densities and $(p,p)$ cross sections,
I calculated the isotopic density differences from $\Ni58$
\begin{align}
&D(\rho_{n};r)\equiv \rho_{n}(\NiA;r)-\rho_{n}(\Ni58;r),\\
&D(\rho_\textrm{ch};r)\equiv \rho_\textrm{ch}(\NiA;r)-\rho_\textrm{ch}(\Ni58;r),
\end{align}
and the isotopic cross section ratios to $\Ni58$
\begin{align}
R(\sigma;\theta_\textrm{c.m.})\equiv \frac{d\sigma(\NiA)/d\Omega}{d\sigma(\Ni58)/d\Omega}.
\end{align}

%%%%%%%%%%%%%%%%%%%%%%%%%%%%%%
%%%%%%%%%%%%%%%%%%%%%%%%%

%%%%%%%%%%%%%%%%%%%%%%%%%%%%%%
\begin{figure}[!htpb]
\includegraphics[width=0.5\textwidth]{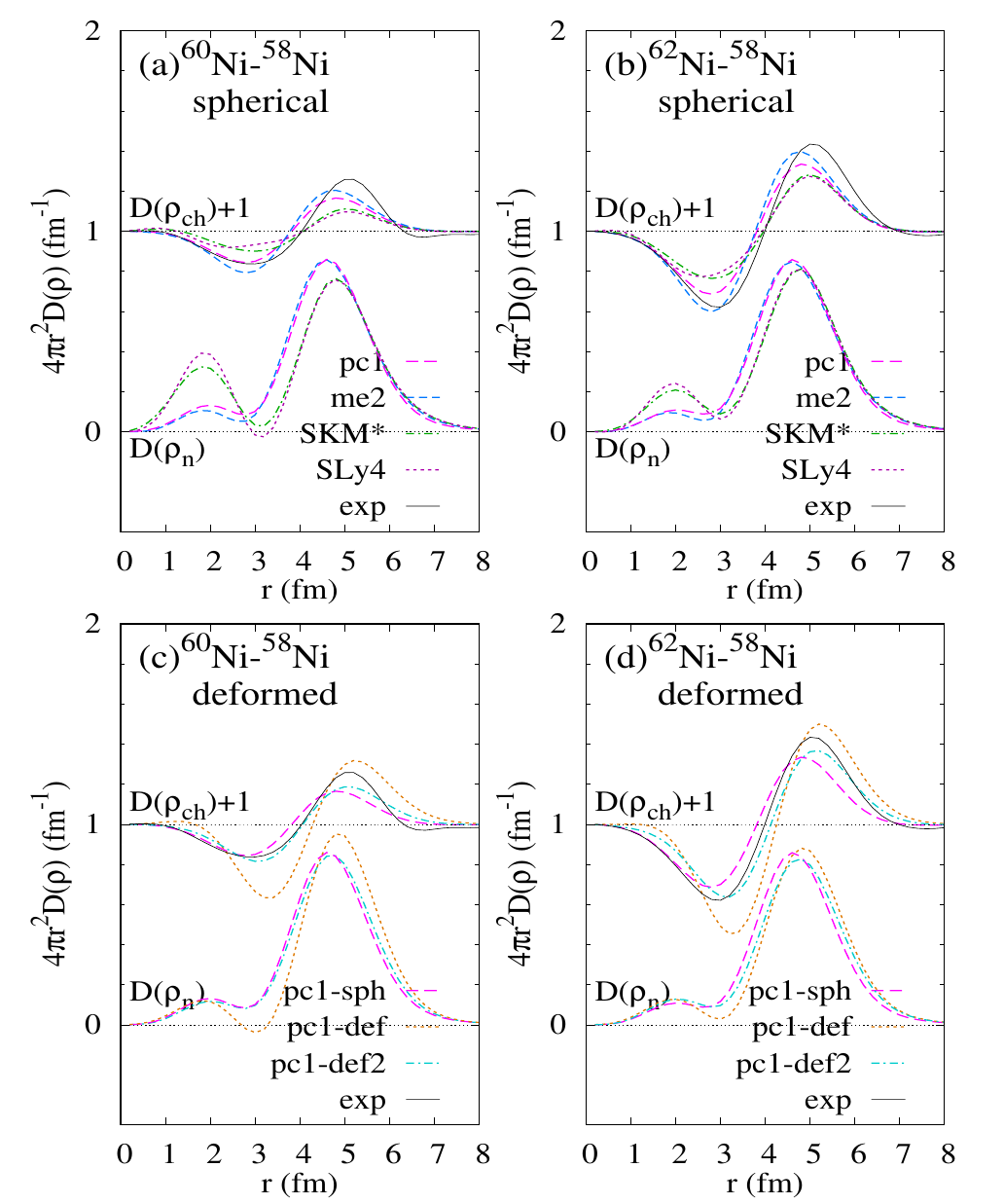}
\caption{
Radial distributions of theoretical
Theoretical neutron density differences and theoretical and
experimental charge density differences in Ni isotopes.
(a) $\Ni60-\Ni58$ and (b) $\Ni62-\Ni58$ differences obtained by the spherical RHB and SHFB calculations, and
(c) $\Ni60-\Ni58$ and (d) $\Ni62-\Ni58$ differences in the pc1-def, pc1-def2, and pc1-sph
states obtained by the deformed RHB calculation.
$4\pi r^2 D_n(r)$ and $4\pi r^2 D_\textrm{ch}(r)+1$ are plotted
The experimental values were
obtained from the SOG fitting parameters listed in Ref.~\cite{DeJager:1987qc}.
\label{fig:dens-compare-ni}}
\end{figure}
%%%%%%%%%%%%%%%%%%%%%%%%%

%%%%%%%%%%%%%%%%%%%%%%%%%%%%%%
\begin{figure}[!htpb]
\includegraphics[width=8.6cm]{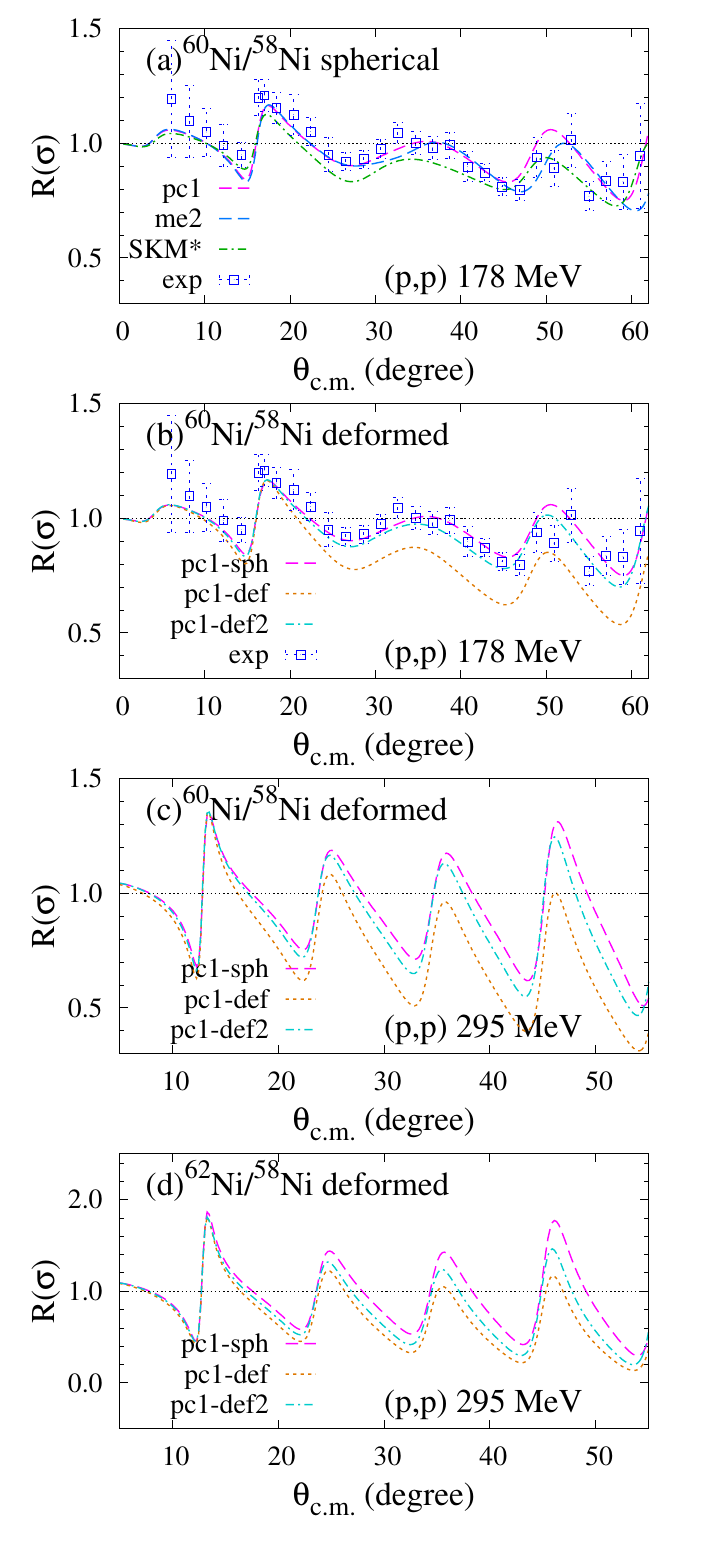}
\caption{
Isotopic cross section ratios $R(\Omega)$ of the $\Ni(p,p)$ reactions at 178~MeV and 295~MeV
obtained by the RIA-ddMH calculations employing the theoretical Ni densities, together with the
experimental $R(\Omega)$ of the $\Ni(p,p)$ reactions at 178~MeV~\cite{Ingemarsson:1981bm}.
(a) $\Ni60/\Ni58$ ratios calculated using the pc1, me2, and SKM* densities obtained by the spherical calculations
at 178~MeV;
$\Ni60/\Ni58$ ratios (b) at 178~MeV and (c) 295~MeV and (d) $\Ni62/\Ni58$ ratios at 295~MeV calculated using
the pc1-def, pc1-def2, and pc1-sph densities obtained by the deformed RHB calculations
The experimental (exp) values are the
$\Ni60/\Ni58$ ratios at 178~MeV~\cite{Ingemarsson:1981bm}.
\label{fig:cross-compare-ni}}
\end{figure}
%%%%%%%%%%%%%%%%%%%%%%%%%

Panels (a) and (b) of ~\ref{fig:dens-compare-ni} display the
$4\pi r^2 D(\rho_n)$ and $4\pi r^2 D(\rho_\textrm{ch})$
obtained by the spherical pc1, me2, SKM*, and SLy4 calculations, together with the experimental
$4\pi r^2 D(\rho_\textrm{ch})$.
The spherical pc1 and me2 calculations slightly underestimate the positions($r$) of the outer peak.
of $4\pi r^2 D(\rho_\textrm{ch})$, whereas
the spherical SKM* and SLy4 calculations significantly underestimate the outer-peak amplitudes.
The different model dependence of $D(\rho_\textrm{ch})$ in the spherical RHB (pc1 and me2) and SHFB (SKM* and SLy4) calculations
can be explained by the neutron single-particle occupations.
Owing to the $N=32$ shell gap (as explained previously), the additional neutrons in $\Ni60$ and $\Ni62$ in the SKM* result [Fig.~\ref{fig:spspe-ni}(d)]
simply occupy the $1p_{3/2}$ orbit and
only weakly change the surface proton densities, whereas
in the pc1 results, the additional neutrons partially
occupy the $0f_{5/2}$ orbit through the pairing effect.
The $0f_{5/2}$ neutron components significantly contribute
to the surface proton densities because they more strongly
affect the $0f_{7/2}$ protons than the $1p_{3/2}$ neutrons.
The different single-particle occupations
between the pc1 and SKM* results are well-clarified in $D(\rho_n)$
[see Figs.~\ref{fig:dens-compare-ni}(a) and (b)].

Let us discuss the deformation effects on the isotopic density differences.
Panels (c) and (d) of Fig.~\ref{fig:dens-compare-ni}
display the results of the deformed (pc1-def and pc1-def2) and spherical (pc1-sph) states.
The pc1-def2 (normal deformation) result of $\Ni58(-0.17)$
shows stronger agreement with the experimental $4\pi r^2 D(\rho_\textrm{ch})$
of both the $\Ni60-\Ni58$ and $\Ni62-\Ni58$ differences than the
pc1-def and pc1-sph results. Meanwhile,
the pc1-def (weak deformation) result of $\Ni58(-0.08)$ largely overestimates
the data. These results suggest that a modest deformation increase around $\beta\sim -0.2$ as
$\Ni58(-0.17)$, $\Ni60(-0.20)$, and $\Ni62(-0.23)$ is likely in the isotopic systematics of Ni isotopes.

The isotopic cross section ratio $R(\sigma)$ is sensitive to the isotopic difference $4\pi r^2D(\rho_n)$
of the surface
neutron densities around $r\sim 4$~fm.
Panels (a) and (b) of Fig.~\ref{fig:dens-compare-ni} compare the
calculated isotopic cross section ratios $R(\sigma)$ of $\Ni60/\Ni58$
at 178~MeV are compared with the experimental values.
The spherical pc1 and me2 calculations reasonably agree with the experimental $R(\sigma)$,
but the spherical SKM* calculation underestimates the ratios.
It again excludes the $N=32$ closure of $\Ni60$.
In the results of the deformed states [Fig.\ref{fig:cross-compare-ni}(b)],
the pc1-def2 result well reproduces the experimental $R(\sigma)$,
whereas the pc1-def result largely underestimates the $R(\sigma)$ amplitudes
because $4\pi r^2D(\rho_n)$ in the $r\sim 4$~fm region is smaller in pc-def than in pc1-def2
as previously described.
Both the pc1-def2 and pc1-sph results reproduce the data within the experimental errors.
The theoretically predicted $\Ni60/\Ni58$ and $\Ni62/\Ni58$ cross section
ratios at 295~MeV are displayed in panels (c) and (d) of
Fig.~\ref{fig:dens-compare-ni}, respectively.
The differences in the 295~MeV cross section ratios
among the pc1-def, pc1-def2, and pc1-sph results exhibit qualitatively
similar trends to the 178~MeV cross section ratios. For further analysis,
high-resolution measurements of $(p,p)$ cross sections of $\Ni58$, $\Ni60$,
and $\Ni62$ are requested.

In the present analysis of the $(p,p)$ reaction,
the drastic deformation change from $\Ni58$ to $\Ni60$ in
the pc1-def set was excluded by the experimental values of the $\Ni60/\Ni58$ cross section ratios
$R(\sigma)$.
Both the pc1-def2 and pc1-sph states yield reasonable $R(\sigma)$
and the better of the two cases
is difficult to conclude from the existing data at 178~MeV.
However, after combining the $D(\rho_\textrm{ch})$ data,
it was concluded that modest deformation changes (pc1-def2:
$\Ni58(-0.17)$, $\Ni60(-0.20)$, and $\Ni62(-0.23)$) best reproduce
both the $D(\rho_\textrm{ch})$ and $R(\sigma)$ data. This result on the
deformations is consistent with the quadrupole deformation parameters
evaluated from the experimental $B(E2)$ values \cite{Pritychenko:2013gwa}.

%\subsubsection{Isotopic analysis of Ni isotopes}

\section{Results of $N=28$ isotones around $\Ti50$} \label{sec:results-ti}

As discussed in the previous sections,
neutron structures such as deformations and single-particle occupations
affect the detailed profile of the surface proton density.
Even in proton magic $N=28$ nuclei,
the neutron deformation affects the surface proton densities
as observed in the isotopic differences $D(\rho_\textrm{ch})$ data
of Ni isotopes.
Performing a similar analysis of the $N=28$ isotones
($\Ca48$ and $\Ti50$), I investigated the effects of proton deformations
on the surface neutron densities and demonstrated
the sensitivity of proton scattering on $\Ti50$ deformation.
The results are presented in the next subsection.

\subsection{Structure properties of $\Ca40$, $\Ca48$, and $\Ti50$}

\begin{figure}[!h]
\includegraphics[width=0.5\textwidth]{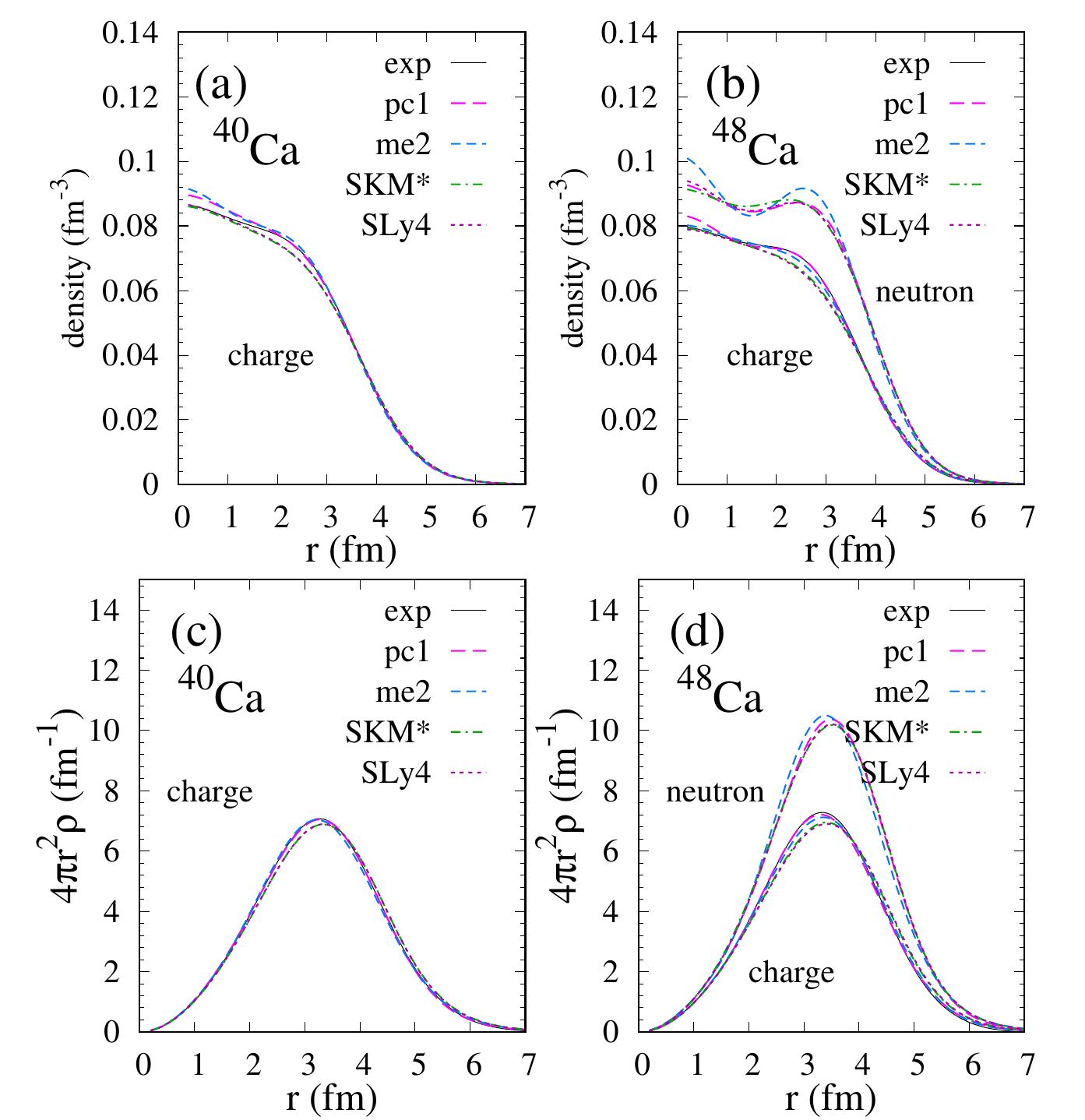}
\caption{Charge $(\rho_\textrm{ch})$ and neutron $(\rho_n)$ densities of Ca isotopes
obtained by the spherical RHB with the pc1 and me2 interactions and by the spherical SHFB with the
SKM* and SLy4 interactions:
(a) $\rho_\textrm{ch}$ of $\Ca40$, (b) $\rho_\textrm{ch}$ and $\rho_n$ of $\Ca48$,
(c) $4\pi r^2 \rho_\textrm{ch}$ of $\Ca40$, and (d) $4\pi r^2 \rho_\textrm{ch}$ and $\rho_n$ of $\Ca48$.
The experimental $\rho_\textrm{ch}$ and $4\pi r^2 \rho_\textrm{ch}$ were
obtained from the SOG fitting parameters listed in Ref.~\cite{DeJager:1987qc}.
\label{fig:dens-ti}}
\end{figure}

The structures of $\Ca40$ and $\Ca48$, and $\Ti50$ were calculated by the spherical and deformed
RHB calculations, and by the spherical SHFB calculations for comparison.
The densities of $\Ca40$ and $\Ca48$ in the spherical case are presented in Fig.~\ref{fig:dens-ti}.
The pc1 and me2 results are consistent with the data, but
the SKM* and Sly4 results fail to reproduce the surface charge density around $r\sim 3$~fm.

%%%%%%%%%%%%%%%%%%%%%%%%%%%%%%
%%%%%%%%%%%%%%%%%%%%%%%%%

%%%%%%%%%%%%%%%%%%%%%%%%%%%%%%
\begin{figure}[!h]
\includegraphics[width=0.5\textwidth]{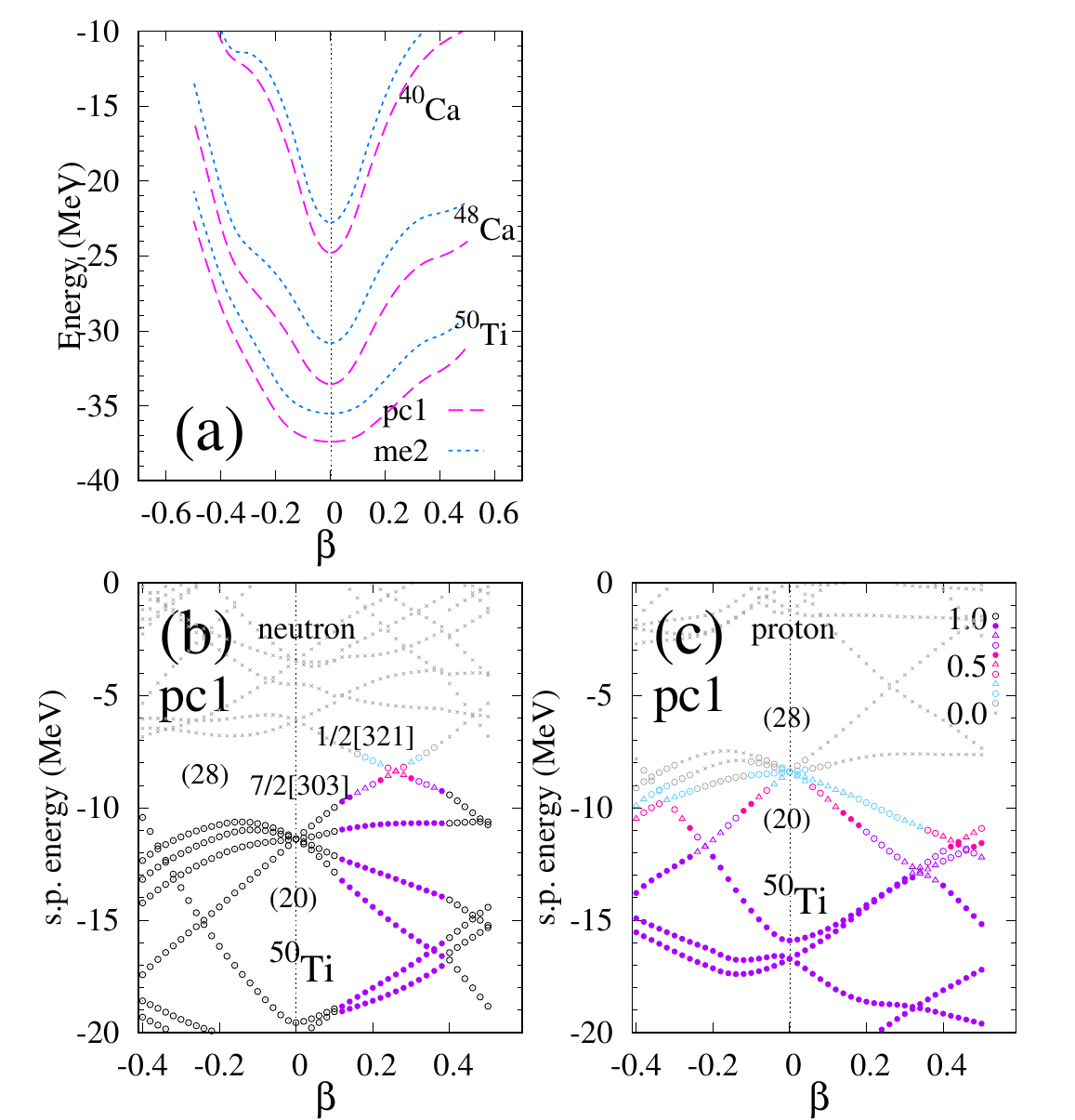}
\caption{
(a) Energy curves of $\Ca40$, $\Ca48$, and $\Ti50$
obtained by the deformed RHB calculations using the pc1 and me2 interactions;
(b) neutron and (c) proton
single-particle energies in $\Ti50$ calculated using the pc1 interactions
(symbol colors indicate the occupation probabilities).
\label{fig:enesp-ti50}}
\end{figure}
%%%%%%%%%%%%%%%%%%%%%%%%%

The $\Ca40$ and $\Ca48$ are double magic nuclei and considered as spherical nuclei, whereas
the ground band of $\Ti50$ is known as a prolate deformation with $\beta\sim 0.16$.
The latter result was determined
from $B(E2)$\cite{Pritychenko:2013gwa} and $\alpha$ scattering~\cite{Rebel:1974mgg}.
The $\beta$-dependent energies obtained by the deformed pc1 and me2 calculations are presented in Fig.~\ref{fig:enesp-ti50}(a),
and those of the neutron and proton single-particle energies of $\Ti50$ are presented
in panels (b) and (c) of Fig.~\ref{fig:enesp-ti50}, respectively.
The energy curves of $\Ca40$ ($\Ca48$) exhibit spherical energy minima
caused by the $Z=20$ and $N=20$($Z=28$) shell effects.
The deformed RHB calculations of $\Ti50$ also yield spherical energy minima,
although the energy curves are soft against deformation.
The spherical $\Ti50$ state is inconsistent with the experimental indication of
prolate deformation.
To discuss the deformation effect of $\Ti50$ on the isotopic systematics of densities and
$(p,p)$ cross sections, I chose the deformed solution $\Ti50(\beta=0.15)$
and compared the results of employing the spherical $\Ti50(\beta=0.0)$ and
deformed $\Ti50(\beta=0.15)$ states.

\subsection{Proton scattering and isotopic analysis of $N=28$ isotones; $\Ca48$ and $\Ti50$}

%%%%%%%%%%%%%%%%%%%%%%%%%%%%%%
\begin{figure}[!htpb]
\includegraphics[width=8 cm]{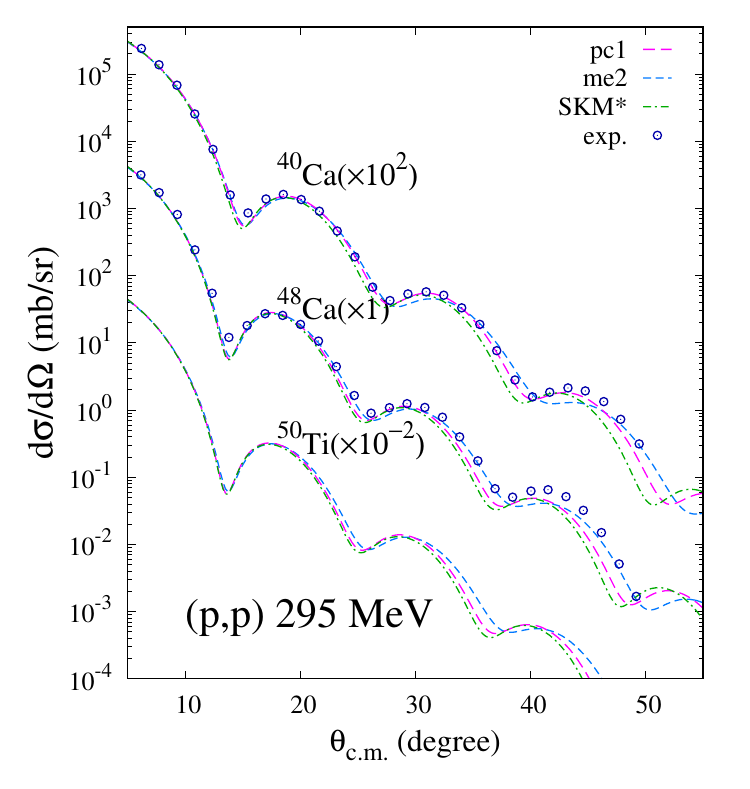}
\caption{
$\Ca40(p,p)$, $\Ca48(p,p)$, and $\Ti50(p,p)$,
cross sections at 295~MeV
calculated by the RIA-ddMH model using the target densities obtained by the spherical pc1, me2, and SKM* calculations.
Also plotted are the experimental data of the 295~MeV
$\Ca40(p,p)$ and $\Ca48(p,p)$ cross sections \cite{Zenihiro:2018rmz}.
\label{fig:cross-ti}}
\end{figure}
%%%%%%%%%%%%%%%%%%%%%%%%%

The $(p,p)$ reactions
of $\Ca40$, $\Ca48$, and $\Ti50$ at 295~MeV
were calculated using the RIA+ddMH model.
Figure~\ref{fig:cross-ti} presents
the calculated cross sections obtained using the
spherical pc1, me2, and SKM* calculations and the experimental data.
The pc1 result agrees with the experimental $\Ca40(p,p)$ cross sections, whereas the
me2 and SKM* results deviate from the data.
For the $\Ca48(p,p)$ cross sections, the pc1 result
reproduces the peak amplitudes of the data with a slight shift of the diffraction pattern
to forward angles, whereas the me2 result underestimates the peak amplitudes
at backward angles. The calculated cross sections in the pc1 and me2 results
differ mainly because the surface neutron densities differ in the two interaction cases.

\begin{figure}[!htpb]
\includegraphics[width=5 cm]{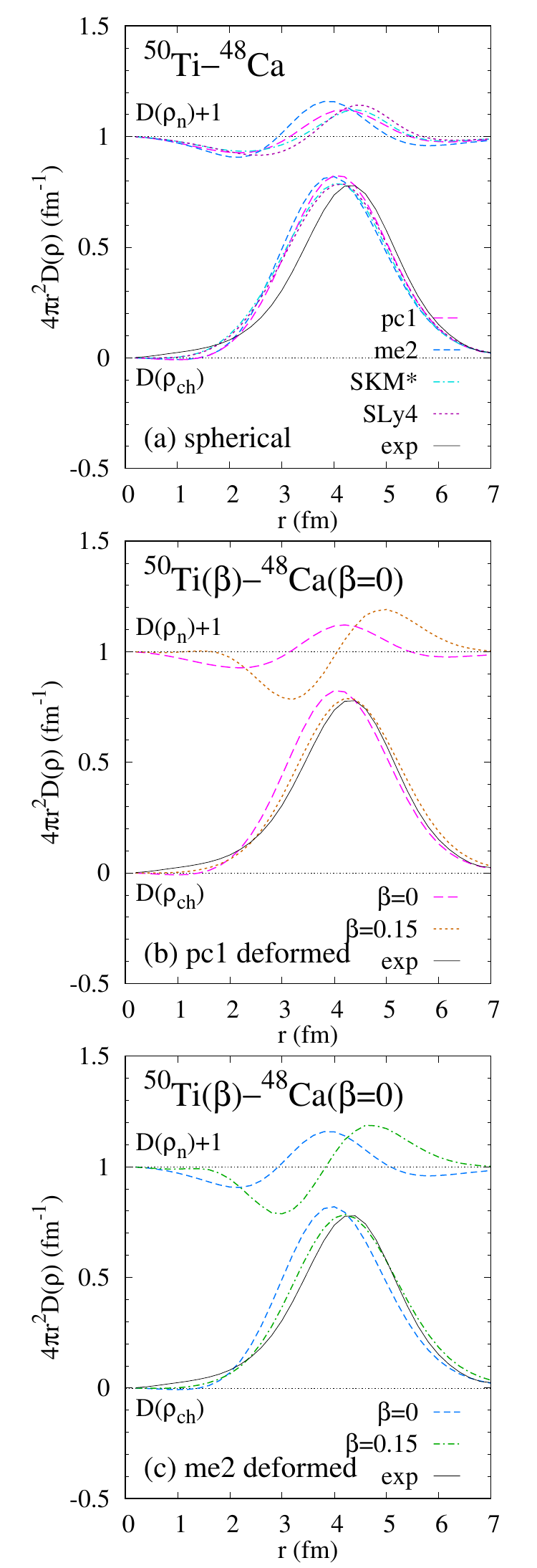}
\caption{
Theoretical neutron and charge density differences between
$\Ti50$ and $\Ca48$, together with the
experimental charge density differences:
(a) density differences obtained by the spherical pc1, me2, SKM*, and SLy4 calculations;
density differences between $\Ti50(\beta=0.15)$ and $\Ca48(\beta=0)$
states obtained by the deformed RHB calculations with the (b) pc1 and (c) me2 interactions,
in comparison with those between $\Ti50(\beta=0)$ and $\Ca48(\beta=0)$.
$4\pi r^2 D_\textrm{ch}(r)$ and $4\pi r^2 D_n(r)+1$ are plotted.
The experimental values were
obtained from the SOG fitting parameters listed in Ref.~\cite{DeJager:1987qc}.
\label{fig:dens-compare-ti}}
\end{figure}

\begin{figure}[!htpb]
\includegraphics[width=8 cm]{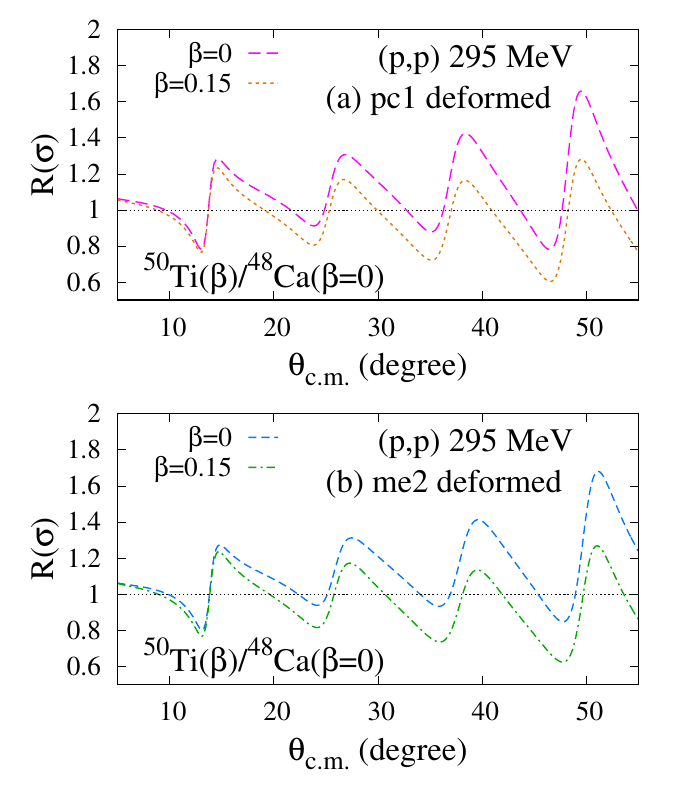}
\caption{
Isotonic $\Ti50/\Ca48$ cross section ratios $R(\Omega)$ of $(p,p)$ reaction at 295~MeV
calculated by the RIA-ddMH model using theoretical densities obtained
by the deformed RHB calculations with the (a) pc1 and (b) me2 interactions.
The results obtained using the
$\Ca48(\beta=0)$ and $\Ti50(\beta=0.15)$ densities are compared with those obtained using the
$\Ca48(\beta=0)$ and $\Ti50(\beta=0.0)$ densities.
\label{fig:cross-compare-ti}}
\end{figure}

To investigate the sensitivity of the $(p,p)$ cross sections
to the detailed surface neutron density profile,
I investigated
the isotonic density differences; between $\Ca48$ and $\Ti50$
\begin{align}
&D(\rho_{n};r)\equiv \rho_{n}(\Ti50;r)-\rho_{n}(\Ca48;r),\\
&D(\rho_\textrm{ch};r)\equiv \rho_\textrm{ch}(\Ti50;r)-\rho_\textrm{ch}(\Ca48;r),
\end{align}
which are presented in Fig.~\ref{fig:dens-compare-ti},
and the isotonic cross section ratios
\begin{align}
R(\sigma;\theta_\textrm{c.m.})\equiv \frac{d\sigma(\Ti50)/d\Omega}{d\sigma(\Ca48)/d\Omega}.
\end{align}
The results are presented in Figs.~\ref{fig:dens-compare-ti}.and ~\ref{fig:cross-compare-ti}, respectively.

%%%%%%%%%%%%%%%%%%%%%%%%%%%%%%
%%%%%%%%%%%%%%%%%%%%%%%%%

%%%%%%%%%%%%%%%%%%%%%%%%%%%%%%
%%%%%%%%%%%%%%%%%%%%%%%%%

None of the spherical calculations can reproduce the
experimentally observed charge density differences $\rho_\textrm{ch}$
[Fig.~\ref{fig:dens-compare-ti}(a)].
In all cases,
the peak position of $4\pi r^2D(\rho_\textrm{ch})$ is shifted inward relative to the
experimental data. The spherical calculation results are significantly improved by
employing the deformed state of $\Ti50(\beta=0.15)$. The outer-region peaks in the $4\pi r^2D(\rho_\textrm{ch})$
profiles of the
$\Ca48(\beta=0.0)$ and $\Ti50(\beta=0.15)$ states are consistent with
the experimental results [Fig.~\ref{fig:dens-compare-ti}(b)].
Under deformation of $\Ti50$ ,
the neutron density difference $D(\rho_n)$ dramatically changes from the spherical
case, reflecting a change in neutron single-particle properties.
As shown in the neutron single-particle energies [Fig.~\ref{fig:enesp-ti50}(c)],
prolate deformation accompanies an energy decrease of
the $1/2[321]$ orbit from the original $1p_{3/2}$ orbit
and quenching of the $N=28$ shell gap.
As a result, the pairing effect imparts a significant $1p_{3/2}$ neutron component to
the deformed state $\Ti50(\beta=0.15)$
This behavior explains the decreased
$4\pi r^2D(\rho_n)$ of the surface neutron densities around
$r=3\sim 4$ fm in Fig.~\ref{fig:dens-compare-ti}(b).

This decrease of $4\pi r^2D(\rho_n)$ at the nuclear surface apparently contributes to
the $\Ti50/\Ca48$ ratios $R(\sigma)$ of the $(p,p)$ cross sections.
Owing to the abovementioned deformation effects on the surface neutron
densities, the predicted $R(\sigma)$ values obtained using the
$\Ti50(\beta=0.15)$ densities
are significantly smaller than those of the
spherical state $\Ti50(\beta=0)$ [Fig.~\ref{fig:cross-compare-ti}(a)].

To check the ambiguity among the structure calculations for $R(\sigma)$ predictions,
I performed similar isotonic analyses of the densities and cross sections in the
spherical $\Ti50(\beta=0.0)$ and deformed $\Ti50(\beta=0.15)$ states obtained by the deformed
me2 calculations.
The me2 results of the density differences and cross section ratios
are presented in panels~\ref{fig:dens-compare-ti}, respectively.
The me2 results are qualitatively
consistent with the pc1 results,
although the $\Ca48$ density and the $\Ca48(p,p)$ cross sections
exhibit different features in the pc1 and me2 calculations
as discussed previously.
This result indicates that the present isotonic analysis can
observe the
$\Ti50$ deformation effect on the $(p,p)$ cross sections
while lowering the model ambiguity of the structure calculations.

\section{Summary}\label{sec:summary}
This study aimed to extract
structure information from the proton elastic scattering off S isotopes
at 320~MeV and Ni isotopes at $E_p=180$~MeV.
To this end, isotopic analyses
were performed by combining the nuclear structure and reaction calculations
and (for comparison) spherical SHFB calculations.
The $(p,p)$ reactions at intermediate energies ($E_p=180\sim 320$~MeV)
were calculated
using the RIA+ddMH model assuming the theoretical densities of the target nuclei.
%The RIA+ddMH calculations using the target densities obtained by the RHB calculations with the
%pc1 interaction reasonably reproduced the $(p,p)$ cross sections for this mass number region.
The RHB calculations with the pc1 interactions
reasonably reproduced the experimental data of the charge densities and $(p,p)$ cross sections in this mass number region, providing detailed analyses of the nuclear structure and $(p,p)$ reactions.
Particular attention was paid to the deformation effects on the isotopic systematics of
the nuclear structures and $(p,p)$ reactions.

The analysis of S isotopes proved that
the surface neutron density can be sensitively probed
by the $(p,p)$ cross sections via isotopic analysis.
The observed isotopic cross section ratios $R(\sigma)$
of 318~MeV proton scattering were better matched by the deformed RHB calculations than by the spherical RHB calculations, confirming that
deformations in $\S32$ and $\S34$ are essential for replicating the experimental data.

In the isotopic analysis of Ni isotopes, modest changes in oblate deformations
in $\Ni58$, $\Ni60$, and $\Ni62$ most acceptably
described the experimental isotopic charge density differences
$D(\rho_\textrm{ch})$ and the cross section ratios $R(\sigma)$
of 178~MeV proton scattering.
These deformations are consistent with the quadrupole deformation parameters
evaluated from the experimental $B(E2)$.

Furthermore, I performed an isotonic analysis of the $N=28$ isotones
$\Ca48$ and $\Ti50$, and discussed the deformation effect of $\Ti50$
on the neutron structure and $(p,p)$ cross sections.
The deformation effect of $\Ti50$ caused a significant decrease in the $\Ti50/\Ca48$ cross section ratios $R(\sigma)$.

Overall, deformation effects are essential for
describing the isotopic systematics of the
experimental $(p,p)$
cross sections.
Combining the structure and reaction calculations in isotopic and isotonic analyses
of the $(p,p)$ cross sections is a useful approach
for extracting nuclear structure information
such as deformations and single-particle features from proton elastic scattering data
at intermediate energies.

\begin{acknowledgments}
This work was supported
by Grants-in-Aid of the Japan Society for the Promotion of Science (Grant Nos. JP18K03617 and 18H05407)
and by a grant of a joint research project of the Research Center for Nuclear Physics at Osaka
University.
\end{acknowledgments}

\bibliography{RIA-S34-refs}

\end{document}